\documentclass[twoside,english,11pt]{article}

\usepackage{geometry,float}
\usepackage{authblk}
\usepackage{natbib}
\usepackage{amsmath}
\usepackage{amssymb}
\usepackage{amstext}
\usepackage{amsthm}
\usepackage{amsfonts}
\usepackage{mathtools}
\usepackage{color}
\usepackage{xcolor}
\usepackage{multirow}
\usepackage{mathrsfs}
\usepackage[hyphens]{url}

\usepackage{algorithm}%
\usepackage{algorithmicx}%
\usepackage{algpseudocode}%
\usepackage{listings}%

\newtheorem{prp}{Proposition}

\newtheorem{rmk}{Remark}
\newtheorem{ass}{Assumption}

\usepackage[hang,small,bf]{caption}
\usepackage[subrefformat=parens]{subcaption}
\DeclareMathOperator*{\argmin}{arg\,min}

\PassOptionsToPackage{hyphens}{url}

\usepackage[hidelinks]{hyperref}
\hypersetup{
	colorlinks,
	linkcolor={red!50!black},
	citecolor={blue!50!black},
	urlcolor={blue!80!black}
}

\mathtoolsset{showonlyrefs=true}

\begin{document}

\title{Wasserstein $k$-Centers Clustering for Distributional Data}
\date{\today}
	
\author[1,3]{Ryo Okano}
\author[2,3]{Masaaki Imaizumi}
	
\affil[1]{Hitotsubashi University}
\affil[2]{The University of Tokyo}
\affil[3]{RIKEN Center for Advanced Intelligence Project}

\maketitle

\vspace{-2em}

\begin{abstract}
    We develop a novel clustering method for distributional data, where each data point is regarded as a probability distribution on the real line. For distributional data, it has been challenging to develop a clustering method that utilizes modes of variation of the data because the space of probability distributions lacks a vector space structure, preventing the application of existing methods devised for functional data. Our clustering method for distributional data takes account of the differences in both means and modes of variation of clusters, in the spirit of the $k$-centers clustering approach proposed for functional data. Specifically, we consider the space of distributions equipped with the Wasserstein metric and define geodesic modes of variation of distributional data using the notion of geodesic principal component analysis. Then, we utilize geodesic modes of clusters to predict the cluster membership of each distribution. We theoretically show the validity of the proposed clustering criterion by studying the probability of correct membership. Through a simulation study and real data application, we demonstrate that the proposed distributional clustering method can improve the quality of the cluster compared to conventional clustering algorithms.  \\
	\noindent%
\end{abstract}

\section{Introduction}
\label{sec:intro}
Cluster analysis is one of the fundamental tools in statistics used to search for homogeneous subgroups of individuals within a data set. Conventional methods, such as the hierarchical clustering \citep{ward1963hierarchical}, $k$-means clustering algorithm \citep{macqueen1967some}, and model-based clustering \citep{banfield1993model}, have been widely used for vector-valued multivariate data. These methods have also been extended to the clustering of more complex data, such as functional data \citep{abraham2003unsupervised, james2003clustering, serban2005cats, chiou2007functional, jacques2013funclust}, and finite-dimensional manifold-valued data \citep{dhillon2001concept, banerjee2005clustering, mardia2022principal}.

This study focuses on distributional data on the real line, which is a type of complex data. Distributional data arises when each data point can be regarded as a probability distribution, and its analysis is gaining increasing attention in statistics and data science \citep{petersen2022modeling, brito2022analysis}. Examples of a distributional data set include age distributions of countries and house price distributions of cities. Because the space of probability distributions does not have a vector space structure, distributional data cannot be analyzed with existing methods devised for multivariate or functional data. Moreover, distributional data cannot be analyzed with methods for finite-dimensional manifold-valued data because the space of probability distributions usually does not have a finite-dimensional tangent space. A common approach for analyzing distributional data is using the geometry provided by the Wasserstein metric in optimal transport \citep{villani2008optimal}. This approach treats each distributional data point as a point in the Wasserstein space \citep{panaretos2020invitation}, a nonlinear metric space of probability distributions equipped with the Wasserstein metric. With this approach, statistical methodologies for analyzing distributional data have recently been developed, particularly on the real line, such as principal component analysis \citep{bigot2017geodesic, cazelles2018geodesic, campbell2022efficient}, regression models \citep{chen2023wasserstein, ghodrati2022distribution, okano2024distribution}, and autoregressive models for distributional time series \citep{zhang2022wasserstein, zhu2023autoregressive}.

Various partitioning clustering methods for distributional data have been proposed. 
In the context of Symbolic Data Analysis \citep{bock2012analysis}, Dynamic Clustering algorithms, which are generalizations of the $k$-means algorithms, for distributional data were proposed by \cite{irpino2006dynamic} and \cite{irpino2014dynamic} using the Wasserstein metric.
They were extended to fuzzy clustering algorithms by \cite{de2015fuzzy} and \cite{irpino2017fuzzy}, and co-clustering algorithms by \cite{de2021co}.
\cite{terada2010non} proposed a $k$-means type algorithm using empirical joint distributions, and \cite{vrac2012copula} presented an algorithm based on the copula analysis. 
\cite{calo2014hierarchical} proposed a method based on a hierarchal mixture modeling of distributional data. 
A fast $k$-means type clustering method based on a modified Wasserstein distance was proposed by \cite{verdinelli2019hybrid}, and 
a robust clustering method based on trimmed $k$-barycenters in the Wasserstein space was proposed by \cite{del2019robust}. 
 \cite{chazal2021clustering} proposed a method for clustering measures via mean measure quantization. 
 \cite{zhuang2022wasserstein} provided evidence for pitfalls (irregularity and non-robustness) of a $k$-means algorithm based on the Wasserstien metric, and established the exact recovery property of a generalization of a $k$-means algorithm for clustering Gaussian measures.
 We note that the above studies consider the clustering of general multivariate distributional data in contrast to this study.

 One challenge in working with distributional data is clustering in a way that incorporates the differences in modes of variation of clusters. While clusters of distributional data usually have different means, they also have different modes or patterns of variation. For example, consider two clusters of distributional data: one cluster consists of population age distributions of districts in a country for men, and the other cluster consists of those for women. These two clusters typically not only have different means, but also different modes of variation. However, this difference in modes of variation may not be captured by the existing clustering methods for distributional data such as the $k$-means method, because they focus only on the differences in means of clusters.
 While \cite{chiou2007functional} developed the $k$-centers clustering for functional data to utilize the modes of variation of clusters, it cannot handle distributional data due to the lack of a formal definition of modes of variation in this setup. Specifically, since distributional data do not have linear forms of basis functions, it is not possible to describe their modes of variation with the ordinary principal component analysis required for $k$-centers clustering.

In this paper, we propose the \textit{$k$-centers distributional clustering}, which takes account of the differences in modes of variation of clusters consisting of distributions on the real line. To this aim, we employ the following steps. First, to define  modes of variation of distributional data, we use the notion of geodesic principal component analysis (geodesic PCA) for distributional data proposed by \cite{bigot2017geodesic}. Specifically, we consider the Wasserstein space of distributions and extract modes of variation of distributional data using geodesics in this space. 
We call the extracted modes as \textit{geodesic modes of variation} of distributional data, and use them to define modes of variation of clusters. 
Next, for every distribution as a data point, we define its predictive models based on the geodesic modes of clusters, and finally determine their cluster memberships by minimizing the discrepancies between the original distributions and their predictive models.

\begin{figure}[h]
\centering
\begin{minipage}[b]{0.4\columnwidth}
    \centering
    \includegraphics[width=0.9\columnwidth]{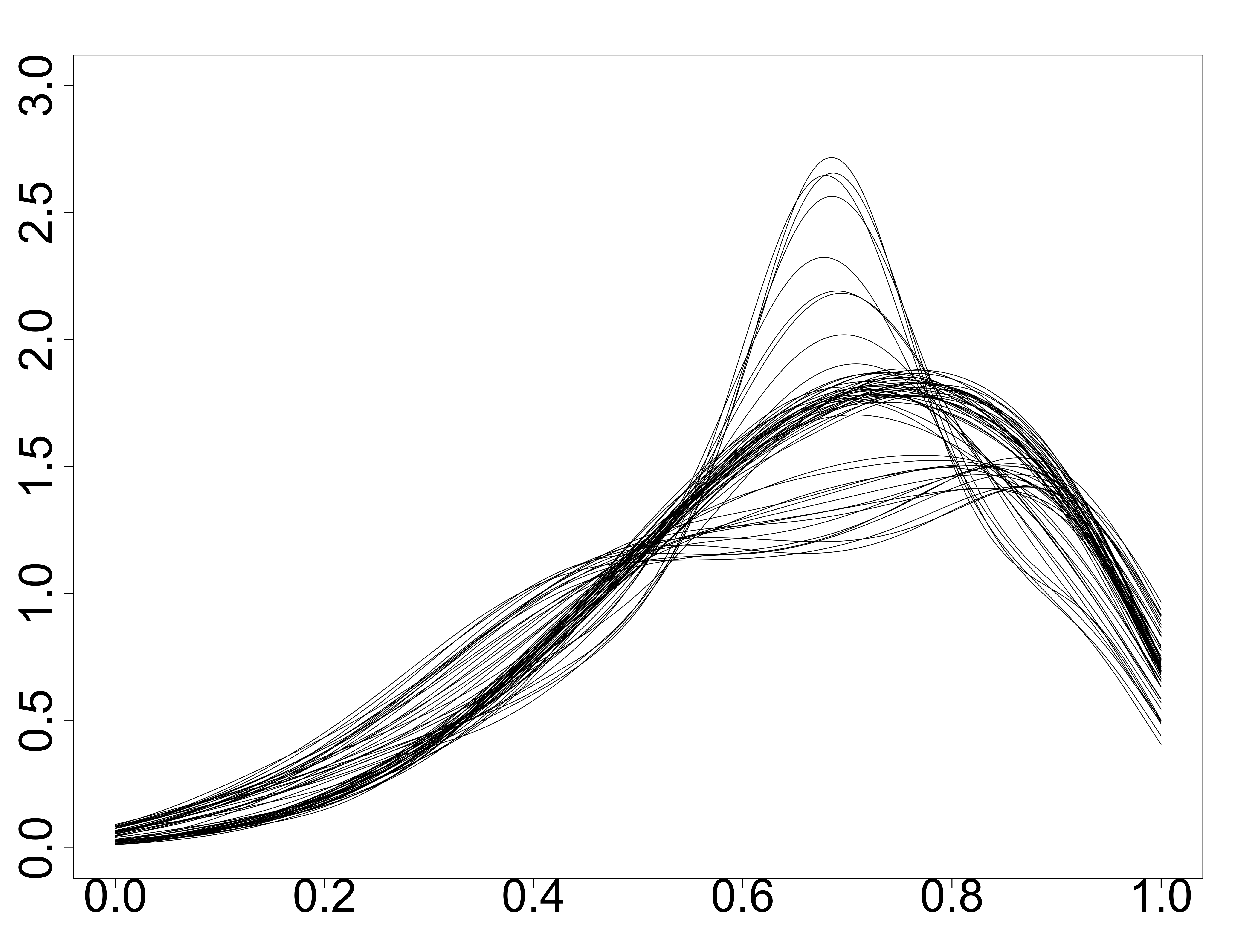}
    \subcaption{Simulated data set}
\end{minipage}\\
\begin{minipage}[b]{0.48\columnwidth}
    \centering
    \includegraphics[width=0.9\columnwidth]{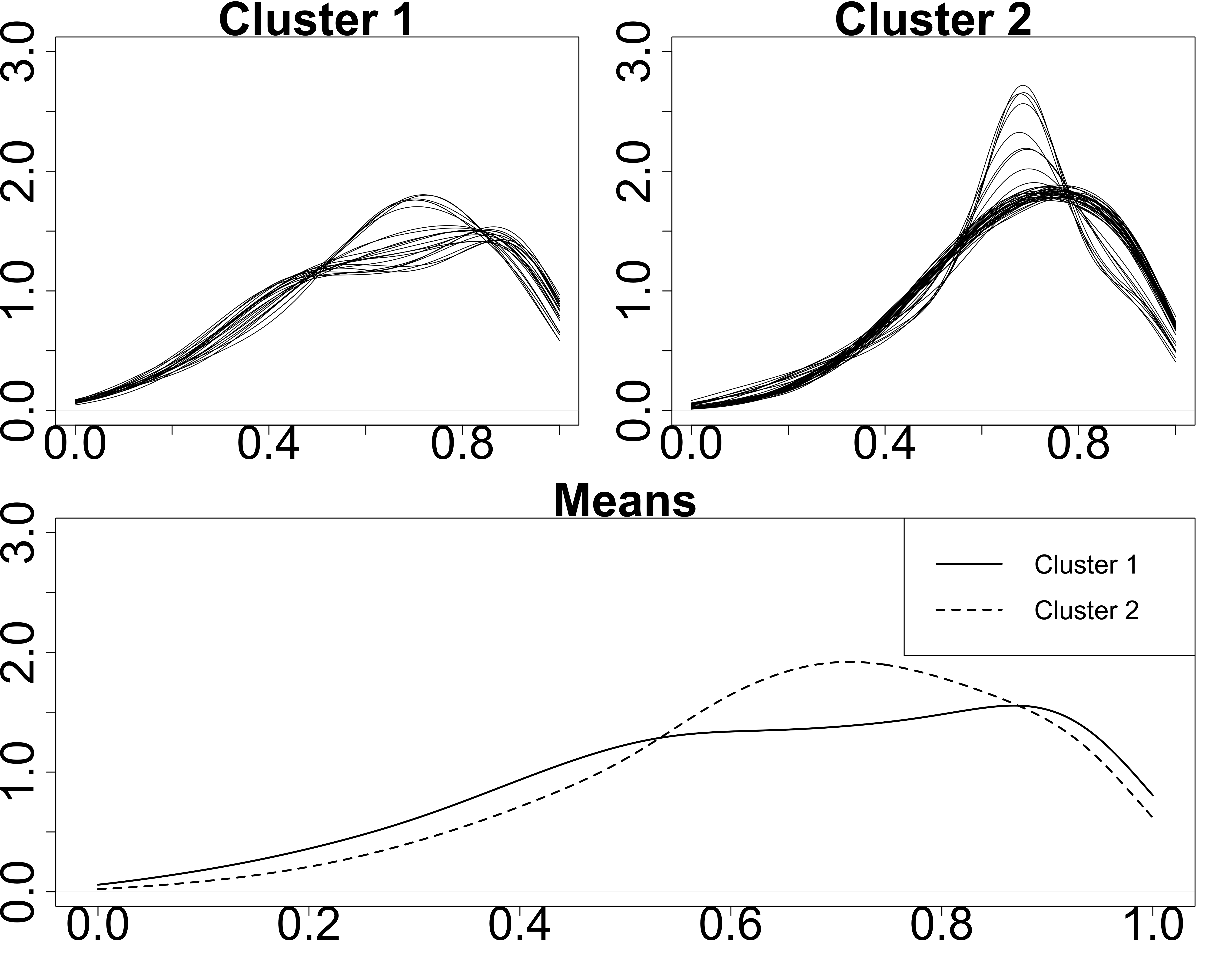}
    \subcaption{Wasserstein $k$-means}
\end{minipage}
\begin{minipage}[b]{0.48\columnwidth}
    \centering
    \includegraphics[width=0.9\columnwidth]{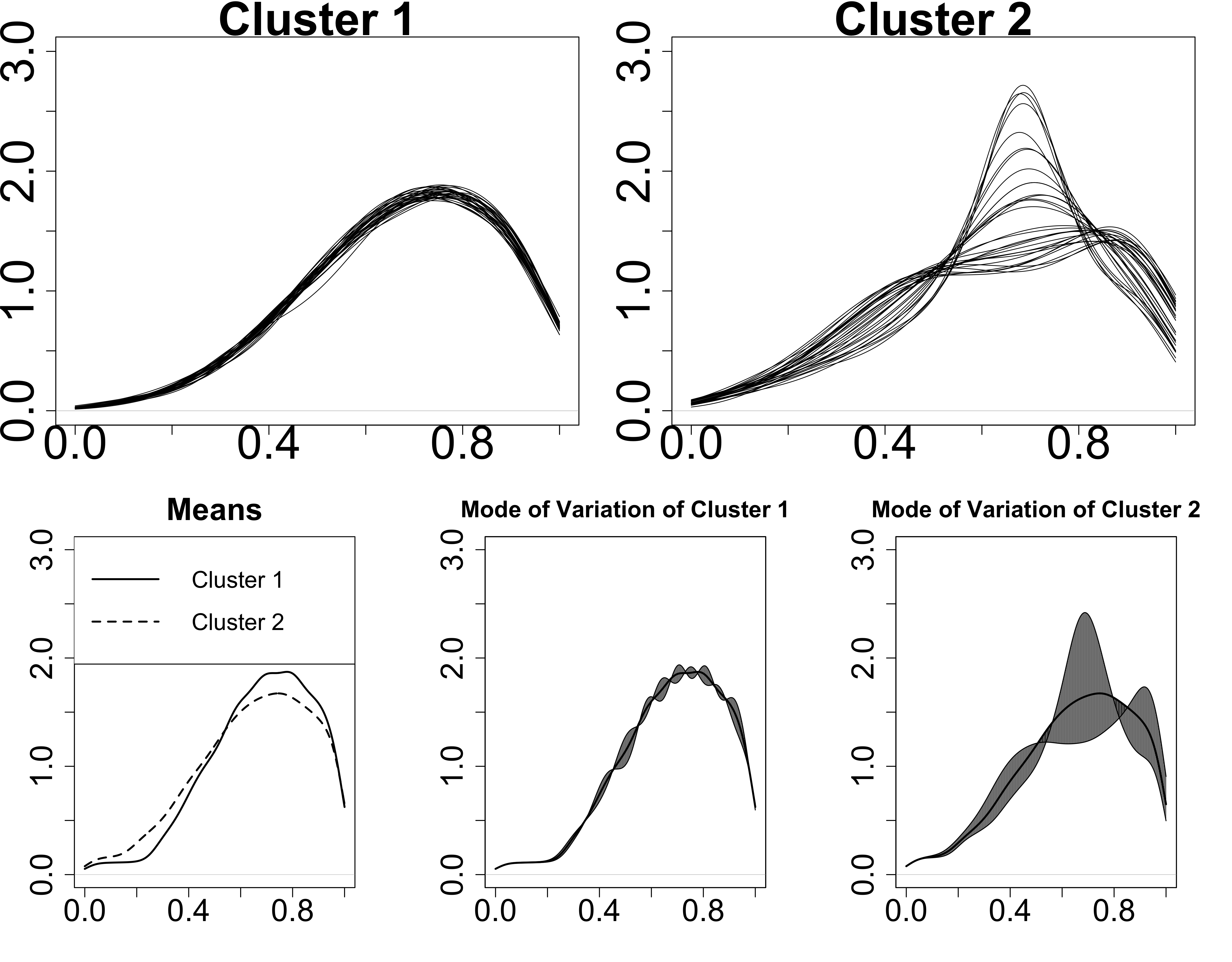}
    \subcaption{Wasserstein $k$-center}
\end{minipage}
\caption{
      Comparison of the Wasserstein $k$-means clustering and the proposed Wasserstein $k$-centers clustering. (a): A simulated data set consisting of sixty distributions on the interval $[0, 1]$. (b): Two partitions obtained by the Wasserstein $k$-means and their Fr\'{e}chet means. (c): Two partitions obtained by the proposed $k$-centers clustering and their Fr\'{e}chet means and modes of variation. The proposed method can detect the two clusters that have similar means and widely different modes of variation.}
      \label{fig:toy_example}
\end{figure}

Our method has several merits. First, the $k$-centers clustering approach can improve cluster quality by considering the difference in modes of variation of clusters, in contrast to the $k$-means approach, which only takes account of the difference in means. This point is demonstrated in our experiment as well as in the work of \cite{chiou2007functional} in the context of functional data clustering. Second, our approach does not rely on any distributional assumptions on data, unlike several model-based clustering approaches that assume a Gaussian model, as seen in \cite{banfield1993model} for multivariate data clustering, and \cite{james2003clustering, jacques2013funclust} for functional data clustering. Third, our method provides visual insight into clusters by exploring their mean and modes of variation structures. Furthermore, our theory validates that the proposed method appropriately controls the probability of a correct assignment of data to a cluster. The experiments confirm these merits.
In Figure \ref{fig:toy_example}, using a simulated data set, we compare the $k$-means clustering based on the Wasserstein distance and the proposed $k$-center clustering.

The remainder of this paper is organized as follows. In Section \ref{sec:background}, we provide background on the Wasserstein space. In Section \ref{sec:method}, we introduce the fundamental principles of the proposed clustering method and present its specific algorithm. In Section \ref{sec:theory}, we theoretically demonstrate the validity of our clustering method. The practical performance of the proposed method is illustrated through a simulation study in Section \ref{sec:simu} and a real data application in Section \ref{sec:real}. We conclude with a brief discussion in Section \ref{sec:discuss}. 
The supplementary material contains more background, implementation details, proof and supplementary results.
The codes implementing the proposed method and data set used in Section \ref{sec:real} are available at \url{https://github.com/RyoOkano21/kCentresDIstributionalClustering}.

\section{Preliminary}\label{sec:background}
\label{subsec:Wspace}
We provide some minimal background on the Wasserstein space needed to present our clustering method. For more details, see, for example, \cite{ambrosio2008gradient, villani2008optimal, bigot2017geodesic, panaretos2020invitation}. 

\subsection{Wasserstein Distance/Space}
Let $\Omega = [a, b]$ be a compact interval in $\mathbb{R}$, and $\mathcal{P}(\Omega)$ be the set of Borel probability measures on $\Omega$.
The 2-Wasserstein distance between $\mu_1, \mu_2 \in \mathcal{P}(\Omega)$ is defined by 
\begin{equation}
 d_W(\mu_1, \mu_2) 
    =
    \left\{\int_0^1 [F_1^{-1}(u) - F_2^{-1}(u)]^2 du \right\}^{1/2}, \notag
\label{eq:wasserstein}
\end{equation}
where $F_1^{-1}$ and $F_2^{-1}$ are the quantile functions of $\mu_1$ and $\mu_2$, respectively. 
It can be shown that the 2-Wasserstein distance $d_W$ is a metric on $\mathcal{P}(\Omega)$,
and the resulting metric space $(\mathcal{P}(\Omega), d_W)$ is called the Wasserstein space of probability distributions. 
For any point $\mu \in \mathcal{P}(\Omega)$ and set $S \subset \mathcal{P}(\Omega)$, we define the distance between them as 
$
d_W(\mu, S)
=
\inf_{\lambda \in S}d_W(\mu, \lambda).
$

\subsection{Tangent Space}
Basic concepts of Riemannian manifolds can be generalized to the Wasserstein space. Let fix a reference measure $\mu_\ast \in \mathcal{P}(\Omega)$, which is assumed to be absolutely continuous with respect to the Lebesgue measure on $\Omega$. 
We define the tangent space of $\mathcal{P}(\Omega)$ at $\mu_\ast$ as the Hilbert space $\mathcal{L}_{\mu_\ast}^2(\Omega)$ of real-valued, $\mu_\ast$-square-integrable functions on $\Omega$,  with an inner product $\langle g_1, g_2 \rangle_{\mu_\ast} = \int_{\Omega}g_1g_2d\mu_\ast$ and a norm $\|g\|_{\mu_\ast} = \langle g, g \rangle_{\mu_\ast}^{1/2}$.
The exponential map $\text{Exp}_{\mu_\ast}: \mathcal{L}_{\mu_\ast}(\Omega) \to \mathcal{P}(\Omega)$ is then defined by 
\begin{equation}
    \text{Exp}_{\mu_\ast}g = (g+\text{id}) \# \mu_\ast, \notag
    \label{eq:expmap}
\end{equation}
where $\text{id}$ denotes the identity function on $\Omega$, and for a measurable function $h: \Omega \to \mathbb{R}$, $h\#\mu_\ast$ is the push-forward measure such that $h\#\mu_\ast(A) = \mu_\ast(h^{-1}(A))$ for any Borel set $A \subset \mathbb{R}$. Moreover, the logarithmic map $\text{Log}_{\mu_\ast}: \mathcal{P}(\Omega) \to \mathcal{L}_{\mu_\ast}^2(\Omega)$ is defined by 
\begin{equation}
    \text{Log}_{\mu_\ast} \mu 
    =
    F^{-1} \circ F_\ast - \text{id},
    \label{eq:logmap}
\end{equation}
where $F_\ast$ and $F^{-1}$ are the distribution function of $\mu_\ast$ and quantile function of $\mu$, respectively. We denote the range of the logarithmic map as $V_{\mu_\ast}(\Omega)$.
The restriction of the exponential map $\mathrm{Exp}_{\mu_\ast}$ to $V_{\mu_\ast}(\Omega)$ is an isometric homeomorphism, the inverse map of which is $\mathrm{Log}_{\mu_\ast}$ (Theorem 2.2, \cite{bigot2017geodesic}). 
Hence, we have
\[
d_W(\mu_1, \mu_2) = \|\mathrm{Log}_{\mu_\ast}\mu_1 - \mathrm{Log}_{\mu_\ast}\mu_2 \|_{\mu_\ast}
\]
for all $\mu_1, \mu_2 \in \mathcal{P}(\Omega)$. 
In addition, the set $V_{\mu_\ast}(\Omega)$ is closed and convex in  $\mathcal{L}_{\mu_\ast}^2(\Omega)$ \citep[Proposition 2.1][]{bigot2017geodesic}.

\begin{rmk}[Connection to the optimal transport map]
\label{rmk:otmap_interpretation}
    From a viewpoint of optimal transport theory, transforming with the logarithmic map is interpreted as  the transformation of a distribution to the corresponding optimal transport map. Specifically, for the probability measures $\mu_\ast$ and $\mu$, any map $T: \Omega \to \Omega$ that minimizes Monge's problem $\inf_{T \# \mu_\ast = \mu} \int_{\Omega} [T(x) - x]^2d\mu_\ast(x)$ is called an optimal transport map from $\mu_\ast$ to $\mu$. In our setting, such optimal transport map uniquely exists and can be expressed as $T = F^{-1} \circ F_\ast$, where $F_\ast$ and $F^{-1}$ are the distribution function of $\mu_\ast$ and quantile function of $\mu$, respectively. Therefore, the function obtained by the logarithmic map \eqref{eq:logmap} is identical to the optimal transport map up to subtraction of the identity function.
\end{rmk}

\subsection{Geodesic}
We explain the notion of geodesic in the Wasserstein space.
A set $G \subset \mathcal{P}(\Omega)$ is called geodesic if for every two points in $G$, there exists a shortest path between them totally contained in $G$.  
A set $G \subset \mathcal{P}(\Omega)$ is geodesic if and only if $\mathrm{Log}_{\mu_\ast}(G)$ is convex in $\mathcal{L}_{\mu_\ast}^2(\Omega)$ \citep[Corollary 2.1][]{bigot2017geodesic}.
For a geodesic set $G \subset \mathcal{P}(\Omega)$, its dimension $\mathrm{dim}(G)$ is defined as the dimension of the convex set $\text{Log}_{\mu_\ast}(G)$ in $\mathcal{L}_{\mu_\ast}^2(\Omega)$, that is, the dimension of the smallest affine subspace of $\mathcal{L}_{\mu_\ast}^2(\Omega)$ containing $\text{Log}_{\mu_\ast}(G)$. We note that
$\mathrm{dim}(G)$ does not depend on the choice of reference measure $\mu_\ast$ \citep[Remark 2.4,][]{bigot2017geodesic}.
If $G$ is a nonempty closed  geodesic set, there exists a unique element $\lambda \in G$ such that $d_W(\mu, G) = d_W(\mu, \lambda)$ by the distance-preserving property of the logarithmic map and the Hilbert projection theorem. 
\section{Method}\label{sec:method}

\subsection{Setting}
Suppose we have $n$ distributions $\nu_1, ..., \nu_n \in \mathcal{P}(\Omega)$ and aim to classify them into $K$ groups. Throughout this paper, we assume the number of clusters $K$ is known or predetermined.
If $K$ is not predetermined, we determine it by performing clustering for different numbers of clusters, and evaluating their performance by some criterion. 
A representative option is using the silhouette method \citep{rousseeuw1987silhouettes}, which can be applied to our method since it can be calculated using the Wasserstein distance between distributions.

\begin{rmk}[Observation of distributional data]
    In practical analysis, the distributions $\nu_i$ are seldom directly observed; instead, we observe independent samples from them. In this case, we need to estimate the distributions from the samples before implementing the clustering method. See Section B.1 in the supplementary material for a description of the estimation procedure.
\end{rmk}

\subsection{Overview and Basic Principle}
\label{sec:basic_principle}
The proposed clustering algorithm involves three steps. In the first step, we select the dimension of geodesics, which is a parameter need to be specified in our method. In the second step, we initialize clusters by applying the $k$-means on an embedding of the data. In the third step, we reclassify each distribution into a new cluster based on the mean and modes of variation of each cluster. This final reclassification step is particularly significant in our algorithm, and we explain its basic principle in the following. 

Let $\{h_i: i=1, ..., n\}$ be given cluster memberships of the $n$ distributions, 
where $h_i \in \{1, ..., K\}$ denotes the label of cluster membership for the $i$-th distribution $\nu_i$. 
In preparation, we capture the mean and modes of variation of each cluster with the notions of the Fr\'{e}chet mean and geodesic PCA in the Wasserstein space \citep{bigot2017geodesic}. Specifically, to capture the mean of cluster $c \in \{1, ..., K\}$, we utilize the unique Fr\'{e}chet mean of the sample in cluster $c$, 
\begin{equation}
    \hat{\nu}_{\oplus}^{(c)}
    =
    \argmin_{\mu \in \mathcal{P}(\Omega)}
    \sum_{i: h_i=c}d_W^2( \nu_i, \mu).  \notag
    \label{eq:frechet_mean}
\end{equation}
To capture the modes of variation of cluster $c$, let define $\mathcal{G}_{c, j}$ for any $j$ as the family of nonempty, closed and geodesic subsets $G \subset \mathcal{P}(\Omega)$, such that $\text{dim}(G) \le j$ and $\hat{\nu}_{\oplus}^{(c)} \in G$. 
Then for a given integer $M \ge 1$, we define a sequence of geodesic subsets $\{\hat{G}_j^{(c)}\}_{j=1}^M$ as a solution of a geodesic PCA problem  
\begin{equation}
    \hat{G}_j^{(c)} \in \argmin_{G \in \mathcal{G}_{c, j}, G \supset G_{j-1}^{(c)}} \sum_{i: h_i=c}d_W^2(\nu_i, G), \notag
            \label{eq:npg}
\end{equation}
for $j=1, ..., M$, with $\hat{G}_0^{(c)} = \{\hat{\nu}_\oplus^{(c)}\}$. 
The set $\hat{G}_M^{(c)}$ is called an $(M, \hat{\nu}_\oplus^{(c)})$-nested principal geodesic of the sample in cluster $c$ (Definition 4.3 in \cite{bigot2017geodesic}), and we utilize it to capture the modes of variation of cluster $c$.
We also call $\hat{G}_M^{(c)}$  as the \textit{geodesic modes of variation} of cluster $c$.

\begin{figure*}[hbt!]
	\centering
	\includegraphics[width=119mm]{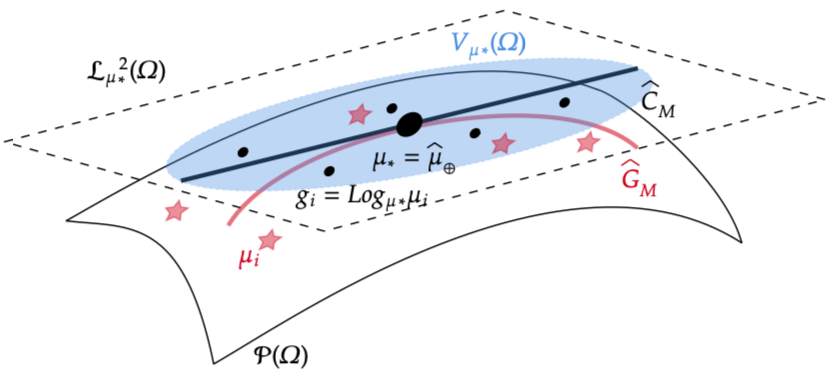}
	\caption{Illustration of the relationship between geodesic PCA in the Wasserstein space and convex PCA in the tangent space (see Section A in the supplementary material for details). 
 The star points are elements $\mu_i$ in the Wasserstein space $\mathcal{P}(\Omega)$, while the the black points are corresponding elements $g_i = \text{Log}_{\mu_\ast}\mu_i$ in the tangent space $\mathcal{L}_{\mu_\ast}^2(\Omega)$. The reference measure $\mu_\ast$ is chosen as the Fr\'{e}chet mean $\hat{\mu}_{\oplus}$ of $\mu_i$'s. The black segment in $\mathcal{L}_{\mu_\ast}^2(\Omega)$ is an $(M, \overline{g})$-nested principal convex component $\hat{C}_M$ of $g_i$'s, and the curve line $\hat{G}_M$ in $\mathcal{P}(\Omega)$ is defined as $\hat{G}_M = \text{Exp}_{\mu_\ast}(\hat{C}_M)$. Then $\hat{G}_M$ is an $(M, \hat{\mu}_\oplus)$-nested principal geodesic of $\mu_i$'s.}
 \label{fig:gpca_illust}
\end{figure*}

\begin{rmk}[Existence of nested principal geodesics and its link to convex PCA]
    In our setting, the existence of an $(M, \hat{\nu}_\oplus^{(c)})$-nested principal geodesic is guaranteed for any $M \ge 1$ (Theorem 4.1, \cite{bigot2017geodesic}). 
    Furthermore, an $(M, \hat{\nu}_\oplus^{(c)})$-nested principal geodesic can be computed by solving a convex principal component analysis (convex PCA) problem in the tangent space. 
We illustrate this relationship in Figure \ref{fig:gpca_illust}.
We give a review of a formulation of convex PCA problem and its link to geodesic PCA problem in Section A in the supplementary material.
\end{rmk}

\begin{figure*}[hbt!]
	\centering
	\includegraphics[width=119mm]{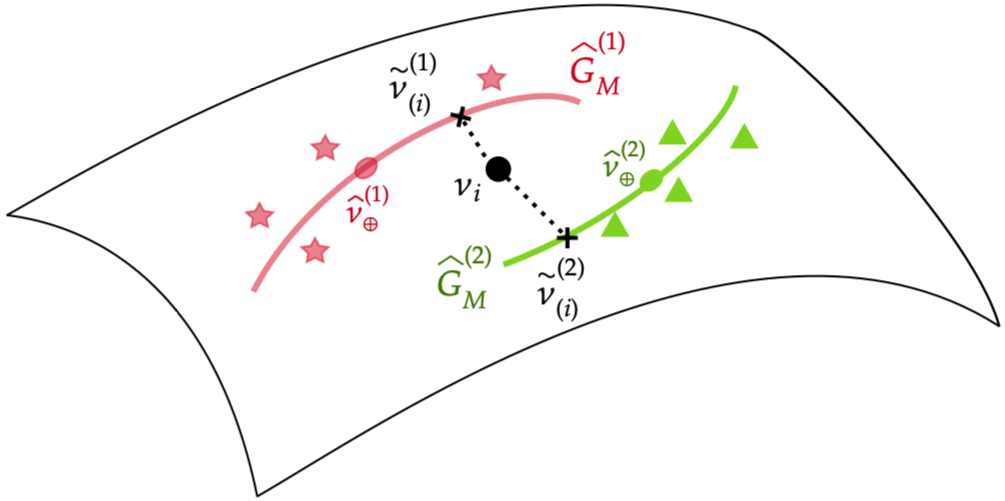}
	\caption{Illustration of the basic principle of the proposed clustering method. For simplicity, we consider the case of two clusters $(K=2)$. The two curves are nested principal geodesics $\hat{G}_M^{(1)}, \hat{G}_M^{(2)}$ as the geodesic modes of the variation, and the star and triangular points around them are samples in clusters $1$ and $2$, respectively. The black point is a data point $\nu_i$, and the cross mark on each curve indicates its model $\tilde{\nu}^{(c)}_{(i)}$. In this case, we have $d_W(\nu_i, \tilde{\nu}^{(1)}_{(i)}) < d_W(\nu_i, \tilde{\nu}^{(2)}_{(i)})$, and thus $\nu_i$ is classified into the cluster 1. }
 \label{fig:basic_principle}
\end{figure*}

Based on the obtained means and modes of variation, we consider to reclassify the $i$-th distribution $\nu_i$ into a new cluster. We define a predictive model of $\nu_i$ as
\begin{equation}
    \Tilde{\nu}^{(c)}_{(i)}
    =
    \argmin_{\mu \in G_M^{(c)}}d_W(\nu_i, \mu),
    \label{eq:pred_model}
\end{equation}
for each $c=1, ..., K$, and reclassify $\nu_i$ by the criterion
\begin{equation}
     \min_{c \in \{1, ..., K\}} d_W(\nu_i, \tilde{\nu}_{(i)}^{(c)}).
    \label{eq:criterion_W}
\end{equation}
We illustrate this principle of our clustering method in Figure \ref{fig:basic_principle}.
In contrast to the $k$-means approach, which only accounts for the difference in means of clusters, the principle of our clustering method accounts for the differences both in the means and modes of variation, potentially improving cluster quality.

\begin{rmk}[Finiteness of dimension $M$]
    We note that the finite-dimensionality of geodesics is essential to our clustering method. By projecting distributional data onto the finite-dimensional principal geodesics of each cluster, clustering can be performed more efficiently, reflecting the information of the clusters, as shown in \citet{chiou2007functional}.
    If we employ infinite-dimensional geodesics to capture the modes of variation of clusters, then the models \eqref{eq:pred_model} can be the perfect approximation of $\nu$ (i.e., $\tilde{\nu}^{(c)}_{(i)} = \nu_i$), and the proposed criterion \eqref{eq:criterion_W} does not work.
\end{rmk}

\begin{rmk}[The dimensions of the geodesic modes]
    In the proposed method, the same value $M$ is used for the dimensions of the geodesic modes of all clusters. This is in contrast to the $k$-centers functional clustering by \cite{chiou2007functional}, where the number of principal components is allowed to vary from cluster to cluster.
    As the dimension of principal geodesic in a cluster $c$ increases, the approximation by the prediction model $\tilde{\nu}^{(c)}$ gets better (i.e., $d_W(\tilde{\nu}^{(c)}_{(i)}, \nu_i)$ approaches to zero) irrelevantly to the mean and modes of variation of the cluster. Therefore, it would be desirable to use the same values of dimension for all clusters for fair comparisons of cluster structures.
\end{rmk}

\begin{rmk}[Clustering of optimal transport maps]
    The proposed clustering for distributional data is similar to the $k$-centers clustering for the transformed functions in the tangent space with the logarithmic map.  
    However, there are also differences.
    One key difference is that we need to explicitly handle the convexity constraint of the set of transformed functions. Because the set of transformed functions (i.e., the range of the logarithmic map $V_{\mu_\ast}(\Omega)$) is a proper convex subset of the tangent space, we need to employ convex PCA introduced by \cite{bigot2017geodesic} to capture their modes of variation. 
    
\end{rmk}

\subsection{Specific Procedure}
\label{sec:procedure}

We describe the specific procedure of the proposed method. Recall that it consists of the three steps: the selection of the dimension of geodesics, initial clustering and reclassification.

\paragraph{Selection of Dimension of Geodesic Modes.}
First of all, we select the dimension $M$ of the geodesics. It is a parameter needed to be specified in the proposed method, and the selected dimension is used in the subsequent steps.
To select it, we define a notion of explained variation for geodesic PCA in the Wasserstein space. Specifically, let $\hat{\nu}_{\oplus}$ be the empirical Fr\'{e}chet mean of the $n$ distributions $\nu_1, ..., \nu_n$ defined by
\begin{equation}
    \hat{\nu}_{\oplus} = \argmin_{\mu \in \mathcal{P}(\Omega)}\frac{1}{n}\sum_{i=1}^n d_W^2(\nu_i, \mu). \notag
    \label{eq:emp_fmean}
\end{equation}
Additionally, for any $M' \in \mathbb{N}$, let $\{\hat{G}_j\}_{j=1}^{M'}$ be a sequence of geodesic subsets defined by  
\begin{equation}
    \hat{G}_{j}
    \in
    \argmin_{G \in \hat{\mathcal{G}}_{j}, G \supset \hat{G}_{j-1}} \frac{1}{n}\sum_{i=1}^n d_W^2(\nu_i, G), \notag
\end{equation}
for $j=1, ..., M'$. 
Here, $\hat{G}_0 = \{\hat{\nu}_\oplus\}$, and  $\hat{\mathcal{G}}_{j}$ is the family of nonempty, closed and geodesic subsets $G \subset \mathcal{P}(\Omega)$, such that $\text{dim}(G) \le j$ and $\hat{\nu}_{\oplus} \in G$. The obtained set $\hat{G}_{M'}$ is an $(M', \hat{\nu}_\oplus)$-nested principal geodesic of the $n$-distributions $\nu_1, ...., \nu_n$.

\begin{rmk}[Computation of the Fr\'{e}chet mean and nested principal geodesic]
When computing the Fr\'{e}chet mean $\hat{\nu}_{\oplus}$ in practice, we use the fact that 
    the quantile function $\hat{F}_{\oplus}^{-1}$ of $\hat{\nu}_{\oplus}$ is expressed as
    \[
    \hat{F}_{\oplus}^{-1}(u)
    =
    \frac{1}{n}\sum_{i=1}^nF_i^{-1}(u), \quad u \in (0, 1),
    \]
    where $F_i^{-1}$ denotes the quantile function of $\nu_i$ (Proposition 4.2, \cite{bigot2017geodesic}).

    Moreover, an $(M', \hat{\nu}_{\oplus})$-nested principal geodesic $\hat{G}_{M'}$ can be computed by solving a convex PCA problem in the tangent space. Specifically, let $\hat{\mu}_\ast \in \mathcal{P}(\Omega)$ be an reference measure, and transform the distributional data $\nu_i$ as 
$g_i = \text{Log}_{\hat{\mu}_\ast}\nu_i$ with the logarithmic map in \eqref{eq:logmap}.
The reference measure $\hat{\mu}_\ast$ is typically chosen as the empirical Fr\'{e}chet mean $\hat{\nu}_{\oplus}$ of $\nu_1, ..., \nu_n$, or the uniform distribution on $\Omega$. 
Then a nested principal convex component  $\hat{C}_{M'} \subset V_{\mu_\ast}(\Omega)$ of $g_1, ..., g_n$ is obtained by solving a convex PCA problem (see Section A in the supplementary material for details), and the set $\hat{G}_{M'} = \rm{Exp}_{\hat{\mu}_\ast}(\hat{C}_{M'})$ is an $(M', \hat{\nu}_{\oplus})$-nested principal geodesic of $\nu_1, ..., \nu_n$ (Proposition 4.4, \cite{bigot2017geodesic}).
\label{rmk:computation}
\end{rmk}

Letting $\tilde{\nu}_i = \argmin_{\mu \in \hat{G}_{M'}}d_W(\nu_i, \mu)$,  the cumulative proportion of variation explained by  $\hat{G}_{M'}$ is defined as
\begin{equation}
    EV(\hat{G}_{M'})
    =
    \frac{n^{-1}\sum_{i=1}^nd_W^2(\tilde{\nu}_{i}, \hat{\nu}_{\oplus})}{n^{-1}\sum_{i=1}^nd_W^2(\nu_i, \hat{\nu}_{\oplus})}.
    \label{eq:cumlative_W}
\end{equation}
We propose to select $M$ as 
\begin{equation}
    M = \min\{M' \in \mathbb{N}: EV(\hat{G}_{M'}) \ge \tau\}, 
    \label{eq:dim_select}
\end{equation}
where $\tau \in (0, 1)$ is a pre-selected threshold value. Setting $\tau = 0.9$ or $\tau = 0.8$ works reasonably well in our numerical experience.

The notion of explained variation for geodesic PCA in the Wasserstein space is equivalent to that for convex PCA in the tangent space, as shown in Section B.2 in the supplementary material . 
We rely on this equivalence when calculating the cumulative proportions of variation \eqref{eq:cumlative_W} in practice. 

\paragraph{Initial Clustering.}
In this step, we classify the $n$ distributions $\nu_1, ..., \nu_n$ into $K$ groups as initialization.
Specifically, we apply the $k$-means algorithm for finite-dimensional vectors obtained by performing dimension reduction on the data.
This approach is analogous to the initial clustering step of the $k$-centers functional clustering method \citep[Section 2.2.1,][]{chiou2007functional}. 

We describe its details.
Let $\hat{\mu}_{\ast} \in \mathcal{P}(\Omega)$ be an absolutely continuous reference measure, and transform the distributional data $\nu_i$ as 
$g_i = \text{Log}_{\hat{\mu}_\ast}\nu_i$ with the logarithmic map in \eqref{eq:logmap}.
Then using the selected $M$ in the previous step, we compute a principal convex component of $g_1, ..., g_n$, and obtain the convex principal component scores of $g_i$, $\hat{\xi}_i = (\hat{\xi}_{i1}, ..., \hat{\xi}_{iM}) \in \mathbb{R}^M$ for each $i=1, ..., n$ (see Section A in the supplementary material for details).
 The initial cluster membership is determined by applying the $k$-means algorithm to the $M$-dimensional vectors $\hat{\xi}_i, i=1, ..., n$.
Let $h_i^{(0)} \in \{1, ..., K\}$ be the label of the cluster membership for the $i$-th distribution $\nu_i$ at this initialization step.

Since the logarithmic map preserve the distance, this initialization is almost equivalent to the Wasserstein $k$-means. A difference is that we first extract $M$ principal components before performing the $k$-means. As we focus on $M$ principal components in the  subsequent reclassification step, this initialization is more suitable for our method, compared to the original Wasserstein $k$-means.

\paragraph{Reclassification.}
With the initial clustering results, we reclassify each distribution into the best predicted cluster with the principle described in Section \ref{sec:basic_principle}. Let $h_i(t) \in \{1, ..., K\}$ be the label of cluster membership for the $i$-th distribution $\nu_i$ at the $t$-th iteration. Given the set of clustering results $\mathcal{H}(t) = \{h_i(t): i=1, ..., n\}$, we obtain for each individual $i$ and cluster $c$  the Fr\'{e}chet mean $\hat{\nu}_{\oplus}^{(c)}$ and an $(M, \hat{\nu}_{\oplus}^{(c)})$-nested principal geodesic 
$\hat{G}_{M}^{(c)}$ based on the distributions $\nu_k$ with $h_k(t) = c$ and $k \neq i$, leaving out the $i$-th  distribution. 
Specifically, we compute the Fr\'{e}chet mean
\begin{equation}
    \hat{\nu}_{\oplus}^{(c)}
    =
    \argmin_{\mu \in \mathcal{P}(\Omega)}
    \sum_{k: h_k(t) = c, k \neq i}
    d_W^2(\nu_k, \mu), \notag
\end{equation}
and a sequence of geodesic subsets $\{\hat{G}_{j}^{(c)}\}_{j=1}^M$ defined by
\begin{equation}
    \hat{G}_{j}^{(c)}
    \in
    \argmin_{G \in \hat{\mathcal{G}}_{c, j}, G \supset \hat{G}_{j-1}^{(c)}} \sum_{k: h_k(t) = c, k \neq i}d_W^2(\nu_k, G), \notag
\end{equation}
for $j=1, ..., M$. Here, $\hat{G}_0^{(c)} = \{\nu_\oplus^{(c)}\}$ and $\hat{\mathcal{G}}_{c, j}$ is the family of nonempty, closed and geodesic subsets $G \subset \mathcal{P}(\Omega)$, such that $\text{dim}(G) \le j$ and $\hat{\nu}_{\oplus}^{(c)} \in G$. 
$M$ is the dimension that was selected according to the criterion \eqref{eq:dim_select} in the first step.
We can compute $\hat{\nu}_{\oplus}^{(c)}$ and $\{\hat{G}_{j}^{(c)}\}_{j=1}^M$ as in Remark \ref{rmk:computation}.

Given these components, we obtain a model of
the distribution $\nu_i$, 
\begin{equation}
    \tilde{\nu}_{(i)}^{(c)}
    =
    \argmin_{\mu \in \hat{G}_M^{(c)}} d_W(\nu_i, \mu)
    \label{eq:pred_model_sample}
\end{equation}
for each cluster $c$ as in \eqref{eq:pred_model}.

In practice, the model $\tilde{\nu}_{(i)}^{(c)}$ can be computed by considering a projection in the tangent space. Specifically, suppose the principal geodesic $\hat{G}_{M}^{(c)}$ is expressed as $\hat{G}_{M}^{(c)} = \text{Exp}_{\hat{\mu}_\ast}(\hat{C}_{M}^{(c)})$ with some reference measure $\hat{\mu}_\ast$ and principal convex component $\hat{C}_{M}^{(c)} \subset V_{\mu_\ast}(\Omega)$. Let transform the distribution $\nu_i$ as $g_i  = \text{Log}_{\hat{\mu}_\ast}\nu_i$ and compute its projection onto $\hat{C}_{M}^{(c)}$, $\tilde{g}_{(i)}^{(c)} = \argmin_{f \in \hat{C}_{M}^{(c)}}\|g_i - f\|_{\hat{\mu}_\ast}$. By the distance-preserving property of the logarithmic map, the model $\tilde{\nu}_i^{(c)}$ is obtained by  $\tilde{\nu}_i^{(c)} = \text{Exp}_{\hat{\mu}_\ast} \tilde{g}_{(i)}^{(c)}$.

The $i$-th distribution $\nu_i$ is then classified into cluster $h_i(t+1)$ such that
\begin{equation}
    h_i(t+1)
    =\argmin_{c \in \{1, ..., K\}} d_W(\nu_i, \tilde{\nu}_{(i)}^{(c)}), 
    \label{eq:criterion_sample}
\end{equation}
as in criterion \eqref{eq:criterion_W}. This step is performed for all $i$, which leads to the updated set of results $\mathcal{H}(t + 1)=\{h_i(t+1): i=1, ..., n\}$. This updating procedure is iteratively implemented until no more data can be reclassified, that is, $\mathcal{H}(t+1) = \mathcal{H}(t)$.

\begin{rmk}
\begin{enumerate}
We describe additional information about the reclustering: 
    \item[(i)]  When reclassifying a data point, we do not use it for the computations of the Fr\'{e}chet mean and principal geodesic of each cluster. 
    This is in order to avoid the problem of overfitting.
    Specifically, suppose a data point $\nu_i$ is in a cluster $c$ which consists of only a few data points. 
    In this case, if we compute the Fr\'{e}chet mean $\hat{\nu}_{\oplus}^{(c)}$ and principal geodesic $\hat{G}_{M}^{(c)}$ using  $\nu_i$ , the model $\tilde{\nu}_{(i)}^{(c)}$ would correspond to the point $\nu_i$ too closely, and the reclassification of $\nu_i$ would be unsuccessful. 
    However, if the cluster $c$ has a sufficient number of data points, the reclassificaiton of $\nu_i$ would work by just computing  $\hat{\nu}_{\oplus}^{(c)}$ and $\hat{G}_{M}^{(c)}$ without leaving out $\nu_i$. This reduces the computational cost of reclassifying $\nu_i$.
    \item[(ii)] There could be an empty cluster similarly as $k$-means. More precisely, if a cluster has only two data points, we cannot compute its principal geodesic to reclassify a point in the cluster. If this situation occurs, we recommend to decrease the number of clusters $K$ and apply the algorithm again.
\end{enumerate}
\end{rmk}

A summary of the proposed clustering procedure is provided in Algorithm
\ref{alg:kcenters}.

\begin{algorithm}
\caption{$k$-Centers Distributional Clustering}
\label{alg:kcenters}
\begin{algorithmic}[1] 
\State Select $K$ as the number of clusters.
\State Select $M$ as the dimension of geodesic modes. 
\State Initialize clusters by applying the $k$-means algorithm for the $M$-dimensional vectors $\hat{\xi}_i$.
\Repeat 
\ForAll {data point $\nu_i$} 
\State{Compute the Frechet mean $\hat{\nu}_{\oplus}^{(c)}$ and an $(M, \hat{\nu}_{\oplus}^{(c)})$-nested principal geodesic $\hat{G}_M^{(c)}$ of every cluster $c$, leaving out the point $\nu_i$.} 
        \State{Compute the model $\tilde{\nu}_{(i)}^{(c)}$ for every cluster $c$.}
        \State{Reclassify $\nu_i$ by the criterion \eqref{eq:criterion_sample}}.
    \EndFor
\Until{no more data point can be reclassified}
\end{algorithmic}
\end{algorithm}

\section{Theory} \label{sec:theory}
In this section, we investigate the theoretical properties of the proposed distributional clustering method. 
Specifically, we demonstrate that the membership criterion \eqref{eq:criterion_W} gives correct clusterings of distributional data.
For full proofs, see Section C in the supplementary material.

\subsection{Setup and Assumptions}
Let assume that there are two clusters labeled $c$ and $d$ in the Wasserstein space $(\mathcal{P}(\Omega), d_W)$, and $\mathcal{P}(\Omega)$-valued random elements $\boldsymbol{\nu}^{(c)}$ and $\boldsymbol{\nu}^{(d)}$ are defined according to the clusters. We define the the Fr\'{e}chet means of $\boldsymbol{\nu}^{(c)}$ and $\boldsymbol{\nu}^{(d)}$ by 
\[
\nu_\oplus^{(c)}
=
\argmin_{\mu \in \mathcal{P}(\Omega)} \mathbb{E}[d_W^2(\boldsymbol{\nu}^{(c)}, \mu)]
\]
and
\[
\nu_\oplus^{(d)}
=
\argmin_{\mu \in \mathcal{P}(\Omega)} \mathbb{E}[d_W^2(\boldsymbol{\nu}^{(d)}, \mu)], 
\] 
respectively. 
 Also, we define $(1, \nu_\oplus^{(c)})$ and $(1, \nu_\oplus^{(d)})$-nested principal geodesics of $\boldsymbol{\nu}^{(c)}$ and $\boldsymbol{\nu}^{(d)}$ by
 \[
 G^{(c)} = \argmin_{G \in \mathcal{G}_c}
 \mathbb{E}[d_W^2(\boldsymbol{\nu}^{(c)}, G)]
 \]
 and
 \[
 G^{(d)}=
 \argmin_{G \in \mathcal{G}_d}
 \mathbb{E}[d_W^2(\boldsymbol{\nu}^{(d)}, G)]
 \]
 respectively, where $\mathcal{G}_c$ is the family of nonempty, closed and one-dimensional geodesic subsets $G \subset \mathcal{P}(\Omega)$ such that $\hat{\nu}_\oplus^{(c)} \in G$, and $\mathcal{G}_d$ is the family of nonempty, closed and one-dimensional geodesic subsets $G \subset \mathcal{P}(\Omega)$ such that $\hat{\nu}_\oplus^{(d)} \in G$.
 Then for the random element $\boldsymbol{\nu}^{(c)}$ of cluster $c$, stochastic models $\tilde{\boldsymbol{\nu}}^{(c)}_c$ and $\tilde{\boldsymbol{\nu}}^{(d)}_c$ are defined as 
\begin{equation}
    \tilde{\boldsymbol{\nu}}^{(c)}_c = \argmin_{\mu \in G^{(c)}} d_W(\boldsymbol{\nu}^{(c)}, \mu), \notag
\end{equation}
and
\begin{equation}
\tilde{\boldsymbol{\nu}}^{(d)}_c = \argmin_{\mu \in G^{(d)}} d_W(\boldsymbol{\nu}^{(c)}, \mu), \notag
\end{equation}
respectively. 

Our theoretical interest is whether the proposed criterion \eqref{eq:criterion_W} correctly determines the cluster membership of $\boldsymbol{\nu}^{(c)}$.
Rigorously, if the condition 
\begin{equation}
    d_W(\boldsymbol{\nu}^{(c)}, \tilde{\boldsymbol{\nu}}^{(c)}_c) <  
    d_W(\boldsymbol{\nu}^{(c)}, \tilde{\boldsymbol{\nu}}^{(d)}_c)
    \label{eq:correct_clustering}
\end{equation}
holds, then the cluster membership of $\boldsymbol{\nu}^{(c)}$ is correctly determined by the proposed criterion \eqref{eq:criterion_W}.
Otherwise, the cluster membership of $\boldsymbol{\nu}^{(c)}$ is wrongly determined by the proposed criterion. 
A goal of our theory is to show that the condition \eqref{eq:correct_clustering} holds with a high probability in several situations.

We state the assumptions. 
Let $\mu_\ast \in \mathcal{P}(\Omega)$ be an absolutely continuous reference measure, and define
$\mathcal{L}_{\mu_\ast}^2(\Omega)$-valued 
random elements $\mathbf{g}^{(c)}$ and $\mathbf{g}^{(d)}$  as $\mathbf{g}^{(c)} = \text{Log}_{\mu_\ast}\boldsymbol{\nu}^{(c)}$ and  
$\mathbf{g}^{(d)} = \text{Log}_{\mu_\ast}\boldsymbol{\nu}^{(d)}$, respectively. 
Then we make the following assumption. 
\begin{ass} 
    Assume that the $\mathcal{L}_{\mu_\ast}^2(\Omega)$-valued random elements $\mathbf{g}^{(c)}$ and $\mathbf{g}^{(d)}$ have expansions
    \begin{equation}
        \mathbf{g}^{(c)}
        =
        m^{(c)}
        +
        \sum_{j=1}^J \xi_j^{(c)} \rho_j^{(c)} \notag 
\end{equation}
and
\begin{equation}
        \mathbf{g}^{(d)}
        =
        m^{(d)}
        +
        \sum_{j=1}^J \xi_j^{(d)} \rho_j^{(d)}, \notag
    \end{equation}
respectively.
Here, $J$ is a positive integer, $m^{(c)}$ and $m^{(d)}$ are some vectors in the range of the logarithmic map $V_{\mu_\ast}(\Omega)$, and $\{\rho_j^{(c)}\}_{j=1}^J$ and $\{\rho_j^{(d)}\}_{j=1}^J$ are some orthonormal vectors in $\mathcal{L}_{\mu_\ast}^2(\Omega)$. 
$\{\xi_j^{(c)}\}_{j=1}^J$ and $\{\xi_j^{(d)}\}_{j=1}^J$ are respectively uncorrelated random variables with zero means such that 
$\mathrm{Var}(\xi_1^{(c)}) 
    \ge  \mathrm{Var}(\xi_2^{(c)}) 
    \ge \cdots 
    \ge
    \mathrm{Var}(\xi_J^{(c)})
\ge0$ 
and 
$
\mathrm{Var}(\xi_1^{(d)}) 
    \ge  \mathrm{Var}(\xi_2^{(d)}) 
    \ge \cdots 
    \ge
    \mathrm{Var}(\xi_J^{(d)})
    \ge
    0
$.
It is also assumed that 
for all $\ell =1, ..., J$, the truncated expansions $m^{(c)} + \sum_{j=1}^\ell \xi_j^{(c)}\rho_j^{(c)}$ and $m^{(d)} + \sum_{j=1}^\ell \xi_j^{(d)}\rho_j^{(d)}$ are in $V_{\mu_\ast}(\Omega)$ with probability equal to 1.
\label{ass:expansion}
\end{ass}

Under Assumption \ref{ass:expansion}, $m^{(c)}$ and $m^{(d)}$ are the mean functions of $\mathbf{g}^{(c)}$ and $\mathbf{g}^{(d)}$, and $\rho_j^{(c)}$ and $\rho_j^{(d)}$ are the $j$-th convex principal directions of $\mathbf{g}^{(c)}$ and $\mathbf{g}^{(d)}$, respectively.
We also assume that the orthogonal projection of $\mathbf{g}^{(c)}$ onto the space $m^{(d)} + \text{span}\{\rho_1^{(d)}\}$ is in $V_{\mu_\ast}(\Omega)$, which is formally stated as follows. 
\begin{ass}
    Assume that the condition 
    \begin{equation}
    m^{(d)} + \langle \mathbf{g}^{(c)} - m^{(d)}, \rho_1^{(d)} \rangle_{\mu_\ast}\rho_1^{(d)} \in V_{\mu_\ast}(\Omega) \notag
    \label{eq:cond_proj}
\end{equation}
holds with probability equal to 1.
\label{ass:inrange}
\end{ass}

\subsection{Correct Membership Probabilities}

In this section, we consider two cases that the two clusters have either the same mean or same covariance, then we study the probability that the condition \eqref{eq:correct_clustering} holds.
Specifically, we express the probabilities by using the random element $\mathbf{g}^{(c)}$ takes values. 
Furthermore, in the subsequent remarks, we evaluate those probabilities more explicitly by imposing a Gaussian distributional assumption on the coefficients of $\mathbf{g}^{(c)}$.

\paragraph{Common Mean Case.}

We consider that the random elements $\boldsymbol{\nu}^{(c)}$ and $\boldsymbol{\nu}^{(d)}$ from the clusters have the same mean structures, i.e., $\nu_{\oplus}^{(c)} = \nu_{\oplus}^{(d)}$ holds. Then, we obtain the following result:
\begin{prp}
    Assume $\nu_{\oplus}^{(c)} = \nu_{\oplus}^{(d)}$ and define a vector $m \in \mathcal{L}_{\mu_\ast}^2(\Omega)$ by $m = \mathrm{Log}_{\mu_\ast}\nu_{\oplus}^{(c)} = \mathrm{Log}_{\mu_\ast}\nu_{\oplus}^{(d)}$. Then under Assumptions \ref{ass:expansion} and  \ref{ass:inrange}, it holds that
    \begin{align}
        &\,\,\,\mathbb{P}(d_W(\boldsymbol{\nu}^{(c)}, \tilde{\boldsymbol{\nu}}^{(c)}_c) <  
    d_W(\boldsymbol{\nu}^{(c)}, \tilde{\boldsymbol{\nu}}^{(d)}_c)) \notag  \\
    &= 
    \mathbb{P}(\langle \mathbf{g}^{(c)} - m, \rho_1^{(c)} + \rho_1^{(d)} \rangle_{\mu_\ast} > 0,  \notag \\ 
    & \qquad \qquad \qquad \langle \mathbf{g}^{(c)} - m, \rho_1^{(c)} - \rho_1^{(d)} \rangle_{\mu_\ast} > 0) \notag  \\ 
    &\,\,\,\,\,\,\,+\mathbb{P}(\langle \mathbf{g}^{(c)} - m, \rho_1^{(c)} + \rho_1^{(d)} \rangle_{\mu_\ast} < 0, \notag \\
    &\qquad \qquad \qquad \langle \mathbf{g}^{(c)} - m, \rho_1^{(c)} - \rho_1^{(d)} \rangle_{\mu_\ast} < 0).
    \label{eq:prob_meansame}
    \end{align}
    \label{prp:prob_1}
\end{prp}

\begin{figure*}
        \centering
        \includegraphics[width=99mm]{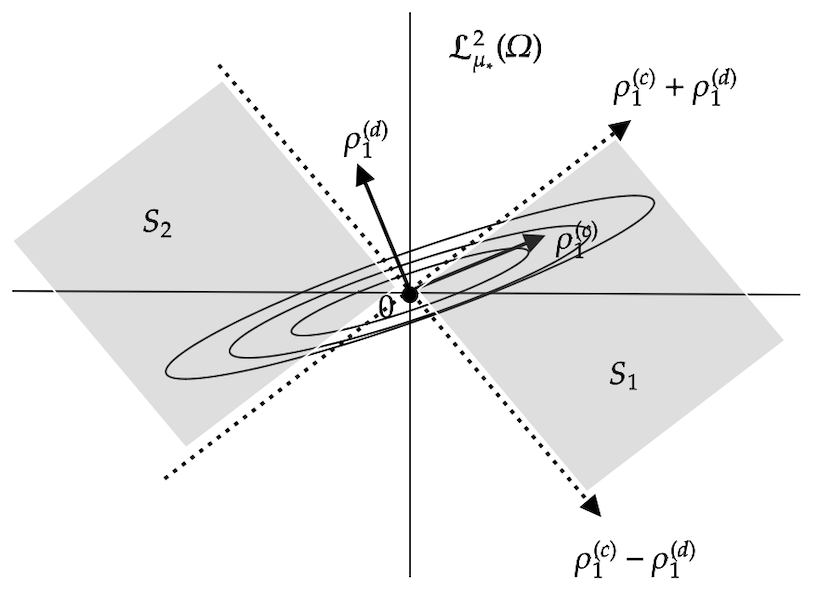}
        \caption{Illustration of the probabilities in Proposition \ref{prp:prob_1}. The contour lines indicate the distribution of the random variable $\mathbf{g}^{(c)}$, and the solid arrows indicate the first convex principal directions $\rho_1^{(c)}$ and $\rho_1^{(d)}$ of $\mathbf{g}^{(c)}$ and $\mathbf{g}^{(d)}$. The dashed arrows indicate the vectors $\rho_1^{(c)} + \rho_1^{(d)}$ and $\rho_1^{(c)} - \rho_1^{(d)}$, and the filled squares indicate the regions $S_1 = \{g \in \mathcal{L}_{\mu_\ast}^2(\Omega): \langle g, \rho_1^{(c)} + \rho_1^{(d)} \rangle_{\mu_\ast} > 0, \langle g, \rho_1^{(c)} - \rho_1^{(d)} \rangle_{\mu_\ast} > 0 \}$ and $S_2 = 
    \{g \in \mathcal{L}_{\mu_\ast}^2(\Omega): \langle g, \rho_1^{(c)} + \rho_1^{(d)} \rangle_{\mu_\ast} < 0, \langle g, \rho_1^{(c)} - \rho_1^{(d)} \rangle_{\mu_\ast} < 0 \}$. As we have $\rho_1^{(c)} \in S_1$ and $-\rho_1^{(c)} \in S_2$, it is expected that $\mathbf{g}^{(c)}-m$ takes values in the regions $S_1$ or $S_2$ with a high probability.}
        \label{fig:iamge_prob}
    \end{figure*}

Proposition \ref{prp:prob_1} implies that when the mean structures of the two clusters are same, 
the probability that the condition \eqref{eq:correct_clustering} holds is equal to the probability that the zero-mean random variable $\mathbf{g}^{(c)} - m$ takes values in regions $S_1$ or $S_2$, where
    \begin{align}
    S_1 = \{g \in \mathcal{L}_{\mu_\ast}^2(\Omega)&: \langle g, \rho_1^{(c)} + \rho_1^{(d)} \rangle_{\mu_\ast} > 0, \notag \\ 
    &\qquad \quad  \langle g, \rho_1^{(c)} - \rho_1^{(d)} \rangle_{\mu_\ast} > 0 \} \notag
    \end{align}
    and
    \begin{align}
    S_2 = 
    \{g \in \mathcal{L}_{\mu_\ast}^2(\Omega)&:
    \langle g, \rho_1^{(c)} + \rho_1^{(d)} \rangle_{\mu_\ast} < 0, \notag \\ 
    &\qquad \quad  \langle g, \rho_1^{(c)} - \rho_1^{(d)} \rangle_{\mu_\ast} < 0 \}, \notag
    \end{align}
    respectively.
    As illustrated in Figure \ref{fig:iamge_prob}, the vectors $\rho_1^{(c)} + \rho_1^{(d)}$ and $\rho_1^{(c)} - \rho_1^{(d)}$ are orthogonal, and if $\rho_1^{(c)} \neq \rho_1^{(d)}$, the first convex principal direction $\rho_1^{(c)}$ of $\mathbf{g}^{(c)}$ satisfies $\rho_1^{(c)} \in S_1$ and $-\rho_1^{(c)} \in S_2$. From these observations, we can expect that the condition \eqref{eq:correct_clustering} holds with a high probability if $\rho_1^{(c)} \neq \rho_1^{(d)}$.

\begin{figure*}    
        \centering
    \begin{minipage}[b]{0.48\hsize}
    \centering
        \includegraphics[width=\hsize]{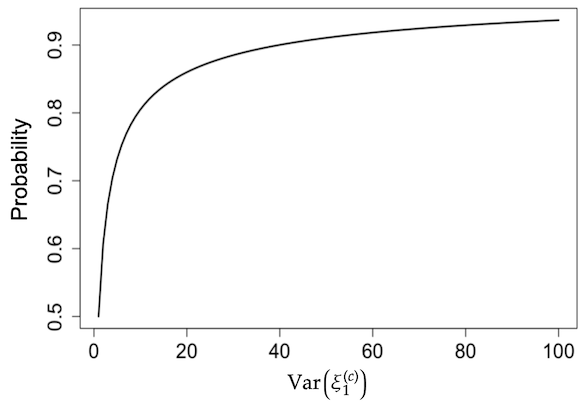}
        \caption{Probability \eqref{eq:prob_meansame_2} as a function of $\mathrm{Var}(\xi_1^{(c)})$. $\mathrm{Var}(\xi_\ell^{(c)})$ is set as $1$. We see that as $\mathrm{Var}(\xi_1^{(c)})$ gets larger, the probability approaches to $1$.}
        \label{fig:prob_gauss_1}
\end{minipage}
\hfill
    \begin{minipage}[b]{0.48\hsize}
    \centering
        \includegraphics[width=\hsize]{Fig5.png}
        \caption{Probability \eqref{eq:prob_gauss_2} as a function of $\|m^{(c)} - m^{(d)}\|_{\mu_\ast}$. $\mathrm{Var}(\xi_\ell^{(c)})$ is set as $1$. We see that as $\|m^{(c)} - m^{(d)}\|_{\mu_\ast}$ gets larger, the probability approaches to 1.}
        \label{fig:prob_gauss_2}
\end{minipage}
    \end{figure*}

    \begin{rmk}
        If we impose a Gaussian distributional assumption on the random coefficients $\xi_1^{(c)}, ..., \xi_J^{(c)}$ of $\mathbf{g}^{(c)}$, we can express the probability \eqref{eq:prob_meansame} more explicitly. Note that since Gaussian distributions have unbounded supports, imposing Gaussianity on the coefficients of $\mathbf{g}^{(c)}$ may violate the assumption that $m^{(c)} + \sum_{j=1}^\ell \xi_j^{(c)} \rho_j^{(c)} \in V_{\mu_\ast}(\Omega)$ holds with probability $1$ for all $\ell = 1, ..., J$. 
        In this remark, 
        we impose a Gaussian assumption instead of the above assumption for the purpose of understanding the probability \eqref{eq:prob_meansame}.
        Specifically, let assume that the $J$-dimensional random vector $(\xi_1^{(c)}, ..., \xi_J^{(c)})$ follows the zero-mean Gaussian distribution with covariance matrix
        \begin{equation}
            \Sigma 
            =
            \begin{pmatrix}
            \mathrm{Var}(\xi_1^{(c)}) & & 0 \\
             & \ddots&  \\
             0 & & \mathrm{Var}(\xi_J^{(c)}) 
\end{pmatrix}. 
\label{eq:gauss_covmat}
        \end{equation}
Then with some additional assumptions, the 
 probability \eqref{eq:prob_meansame} is expressed as 
        \begin{equation}
            0.5 + \frac{1}{\pi}\arcsin\left(\frac{\mathrm{Var}(\xi_1^{(c)}) - \mathrm{Var}(\xi_\ell^{(c)})}{\mathrm{Var}(\xi_1^{(c)}) + \mathrm{Var}(\xi_\ell^{(c)})}\right),
            \label{eq:prob_meansame_2}
        \end{equation}
        with some index $\ell \in \{2, ..., J\}$ (see Section C.2 in the supplementary material for derivation). 
        Since we assumed $\mathrm{Var}(\xi_1^{(c)}) \ge \mathrm{Var}(\xi_\ell^{(c)})$, the value \eqref{eq:prob_meansame_2} is guaranteed to be $0.5$ or higher.
        Also, as illustrated in Figure \ref{fig:prob_gauss_1}, as $\mathrm{Var}(\xi_1^{(c)})$ gets larger,  the value \eqref{eq:prob_meansame_2} approaches to $1$.
        In other words, as the proportion of variation of $\boldsymbol{\nu}^{(c)}$ explained by the principal geodesic $G^{(c)}$ gets larger, the probability that the condition \eqref{eq:correct_clustering} approaches to $1$.
        \label{rmk:gauss_1}
    \end{rmk}

\paragraph{Common Covariance Case.}

We consider that the random elements $\boldsymbol{\nu}^{(c)}$ and $\boldsymbol{\nu}^{(d)}$ from the clusters have the same covariance structures, i.e., 
the $j$-th convex principal directions 
$\rho_j^{(c)}$ and $\rho_j^{(d)}$ are identical 
for $j=1, ..., J$. 
Then, we obtain the following result:
\begin{prp}
    Assume $\rho_j^{(c)} = \rho_j^{(d)}$ holds for $j=1, ..., J$ and denote the common vector by $\rho_j$. 
    Then under Assumptions \ref{ass:expansion} and \ref{ass:inrange}, we have
    \begin{align}
        &\,\,\,\mathbb{P}(d_W(\boldsymbol{\nu}^{(c)}, \tilde{\boldsymbol{\nu}}^{(c)}_c) <  
    d_W(\boldsymbol{\nu}^{(c)}, \tilde{\boldsymbol{\nu}}^{(d)}_c)) \notag \\
    &= \mathbb{P}(\langle \mathbf{g}^{(c)} - m^{(c)}, \psi \rangle_{\mu_\ast} \notag \\ 
    & \qquad \quad > \langle m^{(c)} - m^{(d)}, \rho_1 \rangle_{\mu_\ast}^2 - \|m^{(c)} - m^{(d)}\|_{\mu_\ast}^2),
    \label{eq:prob_covsame}
    \end{align}
    where $\psi = 2m^{(c)} - 2m^{(d)} - 2\langle m^{(c)} - m^{(d)}, \rho_1\rangle_{\mu_\ast}\rho_1$.
    \label{prp:prob_2}
\end{prp}

Proposition \ref{prp:prob_2} implies that when the covariance structures of the two clusters are same,  the probability that the condition \eqref{eq:correct_clustering} holds 
is equal to the probability that the zero-mean random variable $\langle \mathbf{g}^{(c)} - m^{(c)}, \psi \rangle_{\mu_\ast}$ is more than  $\langle m^{(c)} - m^{(d)}, \rho_1 \rangle_{\mu_\ast}^2 - \|m^{(c)} - m^{(d)}\|_{\mu_\ast}^2$.
Since the value $\langle m^{(c)} - m^{(d)}, \rho_1 \rangle_{\mu_\ast}^2 - \|m^{(c)} - m^{(d)}\|_{\mu_\ast}^2$ is non-positive by the Cauchy–Schwarz inequality, we can expect that the condition \eqref{eq:correct_clustering} holds with a high probability.

\begin{rmk}
    As in Remark \ref{rmk:gauss_1}, we can evaluate the probability \eqref{eq:prob_covsame} more explicitly by imposing the Gaussian distributional assumption on the coefficients of $\mathbf{g}^{(c)}$. Let assume the random vector $(\xi_1^{(c)}, ..., \xi_J^{(c)})$ follows the zero-mean Gaussian distribution with covariance matrix \eqref{eq:gauss_covmat}. Then if $J\ge 2$,   the probability \eqref{eq:prob_covsame} is bounded below by
\begin{equation}
         \Phi\left(\frac{\|m^{(c)} - m^{(d)}\|_{\mu_\ast}}{2\sqrt{ \mathrm{Var}(\xi_1^{(c)})}}\right), 
        \label{eq:prob_gauss_2}
    \end{equation}
    where $\Phi$ is the distribution function of the standard normal distribution, and $\mathrm{Var}(\xi_1^{(c)})$ is assumed to be positive (see Section C.2 in the supplementary material for derivation). 
    Since $\|m^{(c)} - m^{(d)}\|_{\mu_\ast} \ge 0$, the value \eqref{eq:prob_gauss_2} is guaranteed to be 0.5 or higher. Also, 
    as illustrated in Figure \ref{fig:prob_gauss_2}, as $\|m^{(c)} - m^{(d)}\|_{\mu_\ast}$ gets larger, the value \eqref{eq:prob_gauss_2} approaches to 1. On the other hand, if $m^{(c)} - m^{(d)} \in \mathrm{span}\{\rho_1\}$, then we have $\tilde{\boldsymbol{\nu}}_{c}^{(c)} = \tilde{\boldsymbol{\nu}}_{c}^{(d)}$ with probability 1. Under this situation, the proposed clustering criterion does not correctly determine the cluster membership of $\boldsymbol{\nu}^{(c)}$. This situation is analogous to a situation called as the non-identifiable situation in \cite{chiou2007functional}, where two models via subspace projection are not distinguished well \cite[Theorem 1,][]{chiou2007functional}.
    \label{rmk:gauss_2}
\end{rmk}

\section{Simulation Study}\label{sec:simu}
In this section, we investigate the practical performance of the proposed method through a simulation study. 

\subsection{Simulation Setup}
We define the following model $\nu_i^{(c)} \in \mathcal{P}(\Omega)$ for each individual $i$ and cluster $c$: 
\begin{equation}
     \nu_i^{(c)} = \text{Exp}_{\mu_\ast} g_i^{(c)}, 
     \label{eq:sim_models}
\end{equation}
where $g_i^{(c)}$ is a random function defined as
\begin{equation}
    g_i^{(c)}(x) = m^{(c)}(x) + 
    \sum_{j=1}^{20} \xi_{ij}^{(c)}\phi_j^{(c)}(x),  \quad x \in \Omega. \notag
\end{equation}
Here, the reference measure $\mu_\ast$ is set to be the uniform measure on $\Omega$, and the mean function $m^{(c)} \in V_{\mu_\ast}(\Omega)$ and the set of principal directions $\mathcal{S}^{(c)} = \{\phi_j^{(c)}\}_{j=1}^{20} \subset \mathcal{L}_{\mu_\ast}^2(\Omega)$ are given by simulation designs. 
For $j=1, ..., 20$, the random coefficient $\xi_{ij}^{(c)}$ is independently generated from the uniform distribution $U[-\lambda_j^{(c)}, \lambda_j^{(c)}]$. The values $\lambda_j^{(c)} (j=1, ..., 20)$ are set such that $\lambda_1^{(c)} \ge \lambda_2^{(c)} \ge \cdots \lambda_{20}^{(c)} > 0$ and $m^{(c)} + \sum_{j=1}^{20}\xi_{ij}^{(c)}\phi_j^{(c)} \in V_{\mu_\ast}(\Omega)$ with probability equal to $1$.

\begin{figure*}
    \centering
    \includegraphics[height=17.5cm]{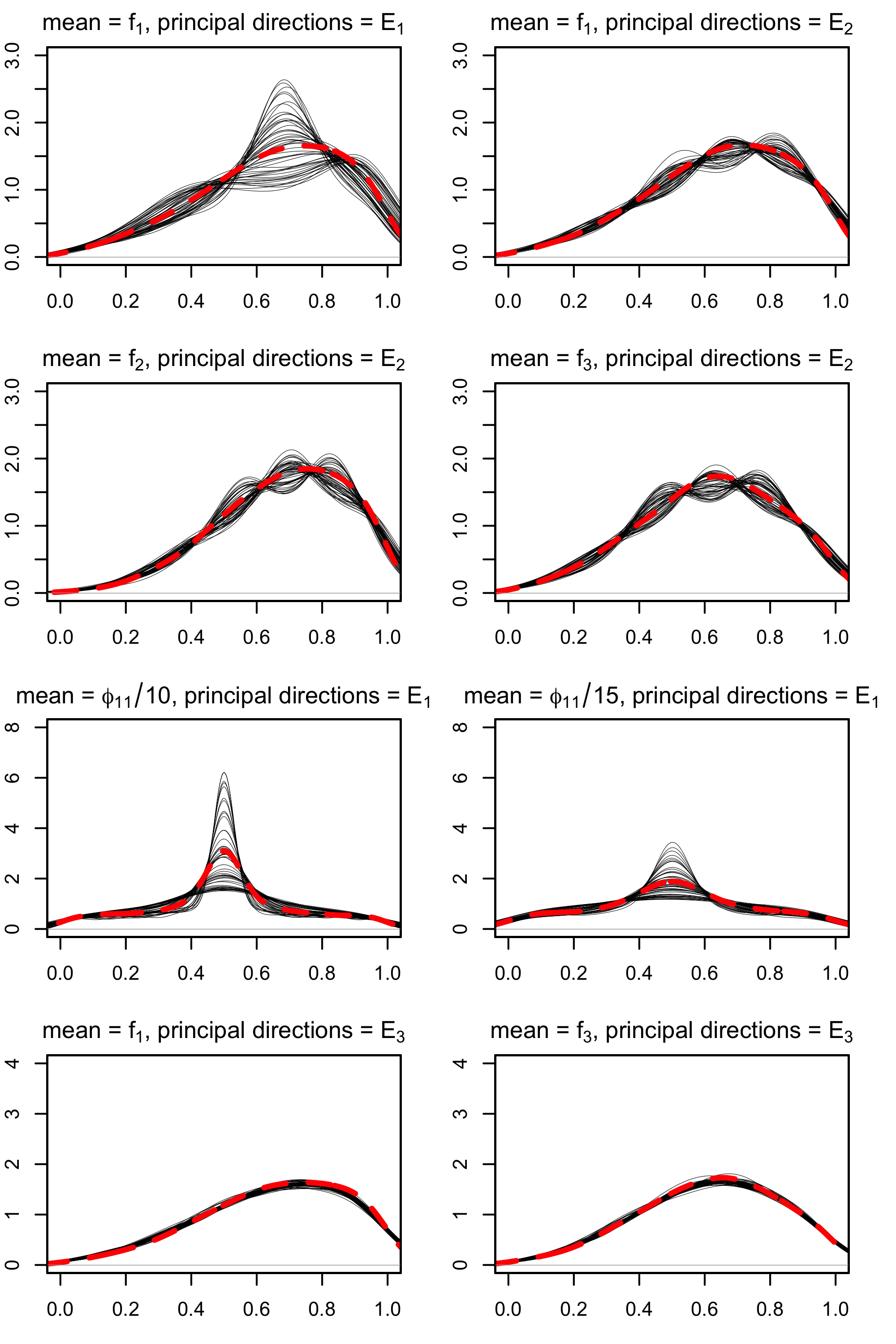}
    \caption{Illustrations of 50 realizations of the random distributions $\nu_i^{(c)}$ with the combinations of the mean function and the set of principal directions. The solid lines are the densities of realizations, and the dotted lines are the densities of the population Fr\'{e}chet means of the clusters.}
    \label{fig:simu_samples_1}
\end{figure*}

We consider various simulation designs with the combinations of $m^{(c)}, \mathcal{S}^{(c)}$ between clusters. They are summarized in Table \ref{tab:simu_designs} with the following notation: 
$f_1(x) = \Phi_{[0, 1]}^{-1}(x; 0.75, 0.3) - x, f_2 = \Phi_{[0, 1]}^{-1}(x; 0.75, 0.25) - x, f_3 = \Phi_{[0, 1]}^{-1}(x; 0.65, 0.25) - x$, where $\Phi_{[0, 1]}^{-1}(\cdot; \mu, \sigma)$ denotes the quantile function of the normal distribution with mean $\mu$ and variance $\sigma^2$ truncated on $[0, 1]$; 
$E_1 = \{\phi_{1j}\}_{j=1}^{20}$, where $\phi_{11}(x) = \sqrt{2}\sin(2\pi x), \phi_{12}(x) = \sqrt{2}\sin(8\pi x)$, $\phi_{1j}(x) = \sqrt{2}\sin((2j+8)\pi x) (j \ge 3)$; $E_2 = \{\phi_{2j}\}_{j=1}^{20}$, where $\phi_{21}(x) = \sqrt{2}\sin(4\pi x), \phi_{22}(x) = \sqrt{2}\sin(6\pi x)$, $\phi_{2j}(x) = \sqrt{2}\sin((2j+8)\pi x) (j \ge 3)$; 
$E_3 = \{\phi_{3j}\}_{j=1}^{20}$, where $\phi_{31}(x) = \sqrt{2}\sin(8\pi x), \phi_{32}(x) = \sqrt{2}\sin(10\pi x)$, $\phi_{3j}(x) = \sqrt{2}\sin((2j+8)\pi x) (j \ge 3)$. 
Except under Design (IV), we set $\lambda_1^{(1)} = 0.4/(2\sqrt{2}\pi), \lambda_2^{(1)} = 0.04/(8\sqrt{2}\pi), \lambda_j^{(1)} = 0.1/\{(2j+8)2^{j-2}\sqrt{2}\pi\}$ for $j \ge 3$; 
$\lambda_1^{(2)} = 0.4/(4\sqrt{2}\pi), \lambda_2^{(2)} = 0.04/(6\sqrt{2}\pi), \lambda_j^{(2)} = 0.1/\{(2j+8)2^{j-2}\sqrt{2}\pi\}$ for $j \ge 3$; 
$\lambda_1^{(3)} = 0.4/(10\sqrt{2}\pi), \lambda_2^{(3)} = 0.04/(12\sqrt{2}\pi), \lambda_j^{(3)} = 0.1/\{(2j+8)2^{j-2}\sqrt{2}\pi\}$ for $j \ge 3$. 
Under Design (IV), we set 
$\lambda_1^{(1)} = \lambda_1^{(2)} = 0.4/(2\sqrt{2}\pi), \lambda_2^{(1)} = \lambda_2^{(2)} = 0.04/(8\sqrt{2}\pi), \lambda_j^{(1)} = \lambda_j^{(2)} = 0.1/\{(2j+8)2^{j-2}\sqrt{2}\pi\}$ for $j \ge 3$. 
In Figure \ref{fig:simu_samples_1}, we illustrate $50$ realizations of the random distributions $\nu_i^{(c)}$ with the combinations of $m^{(c)}$ and $\mathcal{S}^{(c)}$ used in the simulation.
We note  that  Designs (I), (II), (III) and (IV) have 
two clusters while Designs (V), (VI) and (VII) have three clusters. 
Except under Design (IV), the sets of principal directions are different between clusters.
Under Design (VI), it holds that $\text{span} \mathcal{S}^{(1)} \subset \text{span} \mathcal{S}^{(2)}$ and $m^{(1)}, m^{(2)} \in \text{span} \mathcal{S}^{(2)}$. This situation is called the non-identifiable situation by \cite{chiou2007functional}, where it was reported that the $k$-centers functional clustering method was outperformed by other methods. 

\begin{table*}[]
    \centering
    \begin{tabular}{c|cc}
    Design & & \\
     & Mean functions & Sets of principal directions \\ \hline 
     (I) & $m^{(1)} = f_1, m^{(2)} = f_1$ & $\mathcal{S}^{(1)} = E_1, \mathcal{S}^{(2)} = E_2$ \\ 
     (II) & $m^{(1)} = f_1, m^{(2)} = f_2$ & $\mathcal{S}^{(1)} = E_1, \mathcal{S}^{(2)} = E_2$ \\ 
     (III) & $m^{(1)} = f_1, m^{(2)} = f_3$ & $\mathcal{S}^{(1)} = E_1, \mathcal{S}^{(2)} = E_2$ \\ 
     (IV) & $m^{(1)} = \phi_{11}/10, m^{(2)} = \phi_{11}/15$ & $\mathcal{S}^{(1)} = \mathcal{S}^{(2)} = E_1$ \\
     (V) & $m^{(1)} = f_1, m^{(2)} = f_1, m^{(3)} = f_1$ & $\mathcal{S}^{(1)} = E_1, \mathcal{S}^{(2)} = E_2, \mathcal{S}^{(3)} = E_3$ \\ 
     (VI) & $m^{(1)} = f_1, m^{(2)} = f_3, m^{(3)} = f_1$ & $\mathcal{S}^{(1)} = E_1, \mathcal{S}^{(2)} = E_2, \mathcal{S}^{(3)} = E_3$ \\ 
     (VII) & $m^{(1)} = f_1, m^{(2)} = f_2, m^{(3)} = f_3$ & $\mathcal{S}^{(1)} = E_1, \mathcal{S}^{(2)} = E_2, \mathcal{S}^{(3)} = E_3$
    \end{tabular}
    \caption{Simulation designs}
    \label{tab:simu_designs}
\end{table*}

Under every simulation design, we generate a synthetic data set as follows. 
First, we generate distributional data $\nu_i (i=1, ..., 50)$ from the model \eqref{eq:sim_models} with $c=1$, and $\nu_i (i=51, ..., 100)$ from \eqref{eq:sim_models} with $c=2$. If the design is (V), (VI) or (VII), we further generate $\nu_i (i=101, ..., 150)$ from \eqref{eq:sim_models} with $c=3$.
Then for each $i$, we generate 
a set of independent $2000$ samples $Y_i = \{Y_{il}\}_{l=1}^{2000}$ from the generated distribution $\nu_i$.
Our aim is to recover the true cluster memberships based on the synthetic data set $\{Y_i, i=1, ..., n\}$.

\subsection{Other Methods for Comparison, Implementation and Measures of Cluster Quality}

The following methods are applied to the synthetic data set for simulation comparisons: the $k$-means clustering of convex principal component scores (CPCA) that is described in the initial clustering in Section \ref{sec:procedure}; the $k$-centers distributional clustering method (kCDC) which is our proposal; the Wasserstein $k$-means method (WkM) studied by \cite{papayiannis2021clustering} and 
\cite{zhuang2022wasserstein}, for example; 
the $k$-means method with the trimmed Wasserstein distance ($\text{WkM}_{\delta}$) by \cite{verdinelli2019hybrid} with a trimming constant $\delta\in \{0.01, 0.05, 0.1\}$. 

For selecting the dimension $M$ of geodesic modes, we set the threshold value $\tau=0.9$. In the initial clustering and reclassification steps, we use a computational algorithm developed by \cite{campbell2022efficient} for implementations of convex and geodesic PCA. 

To compare performance of the clustering methods, we use two measures of cluster quality. The first is the correct classification rate (\textit{cRate}), which is defined as the maximal possible ratio of correctly classified objects to the total number of objects to be clustered. By definition, \textit{cRate} takes a value between $0$ and $1$, and a larger \textit{cRate} indicates a better clustering quality. 
The second measure of clustering quality is the adjusted Rand index (\textit{aRand}) \citep{hubert1985comparing}, which is a corrected form of the Rand index \citep{rand1971objective}. The Rand index measures the agreement between two partitions by counting the number of paired objects that are either in the same group or in different groups in both partitions. If the two partitions are an external criterion and a clustering result, then the Rand index can be viewed as the quality of the clustering. 
\textit{aRand} is a form of the Rand index that has an expected value 0 and is bounded above by 1. As with \textit{cRate}, a larger \textit{aRand} indicates a better clustering quality.
The index \textit{aRand} is widely used as a measure of clustering quality \citep[for example,][]{chiou2007functional, chiou2008correlation, golzy2020poisson, wei2024skeleton}.

\subsection{Result}

Table \ref{tab:simu_results} summarizes simulation results. 
For each design, the averages of $100$ replications and their standard errors are reported.
In Designs (I), (II), (V) and (VII), the averaged outcomes of the proposed method are much better than those of the other methods.  
In Designs (III) and (VI), the averaged outcomes of the all methods are comparable with each other, but in Design (III) the standard deviation of the proposed method is smaller than those of the other methods.
When the clusters are in the non-identifiable situation in Design (VII), the proposed method is outperformed by the other methods. 
In summary, except under the non-identifiable situation of Design (IV), the proposed method shows the best performances.

\begin{table*}[]
    \centering
    \begin{tabular}{c|cccccc}
        Design&     &     &  \\ 
              & CPCA & kCDC & WkM  & $\text{WkM}_{0.01}$ & $\text{WkM}_{0.05}$ & $\text{WkM}_{0.1}$ \\ \hline
        (I)   &     &      &  \\
        cRate &  0.701 (0.032) & \textbf{0.783} (0.062) & 0.701 (0.033) & 0.701 (0.033) & 0.701 (0.033) & 0.701 (0.033) \\
        aRand &  0.161 (0.051) & \textbf{0.328} (0.156) & 0.161 (0.051) & 0.161 (0.051) & 0.161 (0.051) & 0.161 (0.051) \\
         (II) &     &     &  \\
        cRate & 0.795 (0.043) & \textbf{0.871} (0.056) & 0.795 (0.044) & 0.795 (0.044) & 0.795 (0.043) & 0.770 (0.042) \\
        aRand & 0.350 (0.103) & \textbf{0.557} (0.175) & 0.351 (0.105) & 0.352 (0.105) & 0.349 (0.103) & 0.293 (0.093) \\
        (III) &     &     &  \\
        cRate & 0.969 (0.071) & 0.974 (0.041) & 0.969 (0.071) & 0.969 (0.071) & 0.969 (0.071) & \textbf{0.979} (0.059)  \\ 
        aRand & 0.900 (0.227) & 0.904 (0.146) & 0.897 (0.227) & 0.897 (0.228) & 0.900 (0.226) & \textbf{0.933} (0.189) \\
         (IV) &       \\
        cRate & 0.683 (0.045) & 0.616 (0.055) & \textbf{0.684} (0.044) & \textbf{0.684} (0.045) & \textbf{0.684} (0.044) & 0.683 (0.045) \\ 
        aRand & 0.134 (0.065) & 0.057 (0.053) & \textbf{0.135} (0.065) & \textbf{0.135} (0.067) & \textbf{0.135} (0.065) & 0.134 (0.066) \\ 
        (V)   &     &     &  \\
        cRate & 0.552 (0.028) & \textbf{0.625} (0.048) & 0.553 (0.028) & 0.554 (0.028) & 0.554 (0.028) & 0.556 (0.025) \\ 
        aRand & 0.177 (0.046) & \textbf{0.292} (0.068) & 0.178 (0.047) & 0.179 (0.047) & 0.179 (0.047) & 0.181 (0.044) \\
        (VI)  &     &     &  \\
        cRate & 0.800 (0.021) & \textbf{0.804} (0.037) & 0.801 (0.021) & 0.800 (0.021) & 0.800 (0.021) & 0.800 (0.021) \\ 
        aRand & 0.602 (0.019) & {0.576} (0.052) & \textbf{0.602} (0.019) & \textbf{0.602} (0.019) & \textbf{0.602} (0.019) & 0.601 (0.019) \\
        (VII) &     &     &  \\
        cRate & 0.860 (0.029) & \textbf{0.958} (0.037) & 0.863 (0.029) & 0.863 (0.029) & 0.863 (0.029) & 0.847 (0.028) \\ 
        aRand &  0.677 (0.046) & \textbf{0.884} (0.093) & 0.682 (0.046) & 0.682 (0.046) & 0.682 (0.047) & 0.658 (0.042)
    \end{tabular}
    \caption{Simulation results. For each design, the averages of $100$ replications and their standard errors are reported. 
    The greater the index cRate and cRand, the higher the quality of the clustering. The proposed method is $k$-centers distributional clustering (kCDC). The best results are in bold.}
    \label{tab:simu_results}
\end{table*}

\begin{figure*}[]
    \centering
    \includegraphics[width=0.75\linewidth]{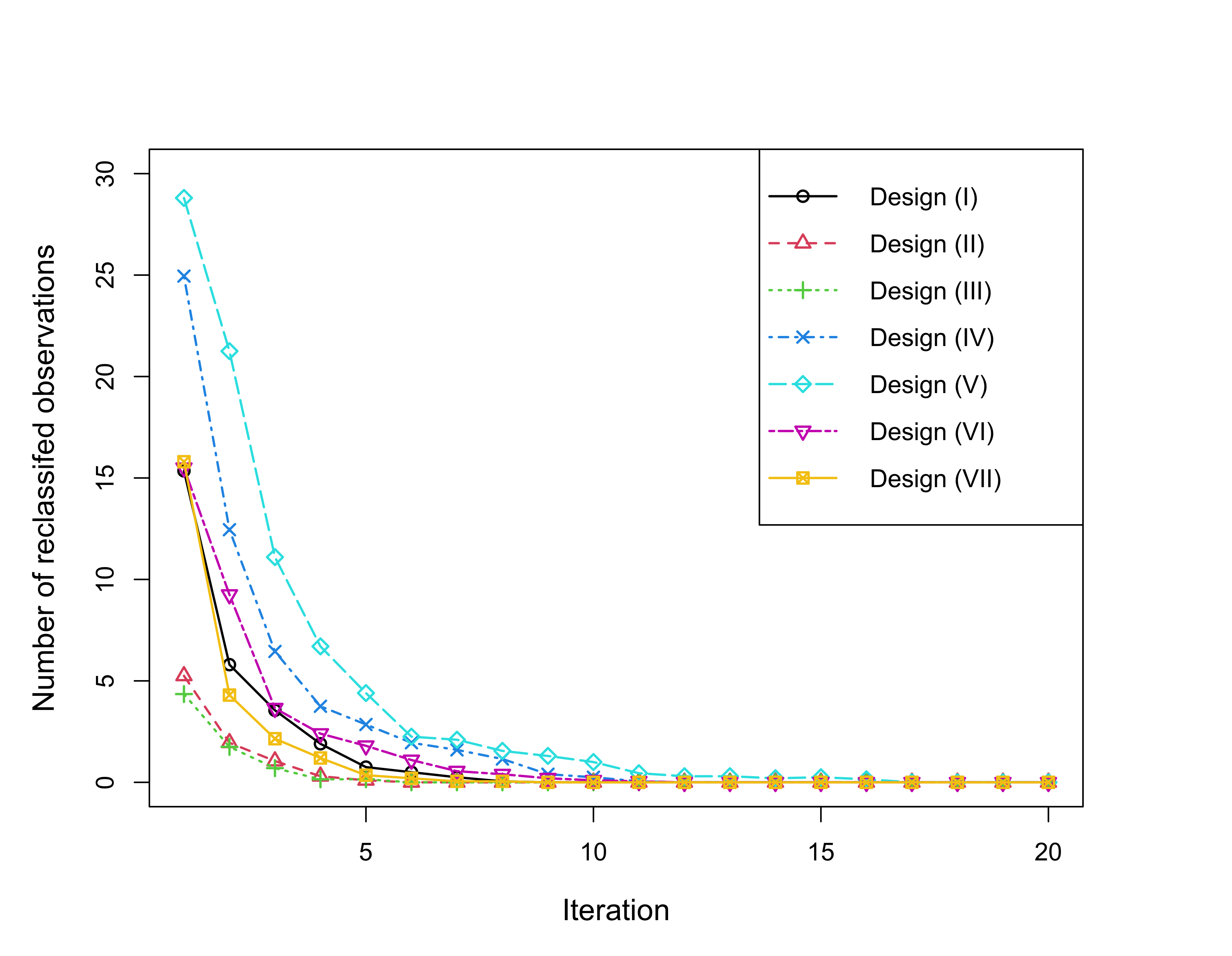}
    \caption{The numbers of distributions reclassified at iteration $t=1, ..., 20$ in the proposed method  under the seven simulation designs. 
    The average of 100 replications is reported for each design.}
    \label{fig:convergence_result}
\end{figure*}

In addition, Figure \ref{fig:convergence_result} shows the averaged number of reclassified distributions in the proposed method. Under all designs, the averaged numbers of reclassified distributions reach zero, and hence the updates stop within 20 iterations.

\section{Real Data Analysis}\label{sec:real}

\subsection{Data}
\begin{figure*}
    \centering
    \includegraphics[width=119mm]{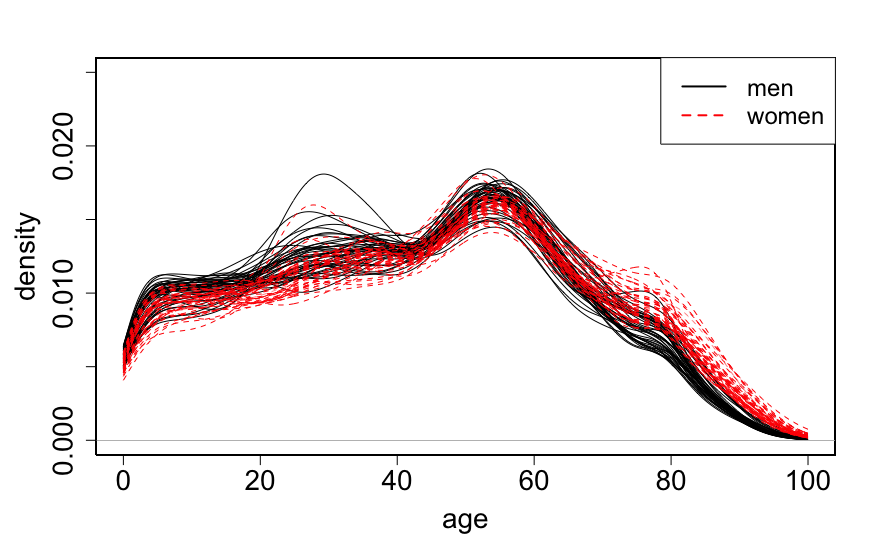}
    \caption{Densities of population age distributions for men and women in the 42 districts in Upper and Lower Austria for the year 2020. The black solid lines are for men, and the red dashed lines are for women.}
    \label{fig:austria_2020_densities}
\end{figure*}

We use a real dataset consisting of population age distributions of districts in Austria for comparing results of several clustering methods. 
The raw data are available from STATcube – Statistical Database of Statistics Austria (\url{https://www.statistik.at/en/databases/statcube-statistical-database}).
For a given year, district and sex, this database provides a cross-sectional table in which the number of people at each age is recorded. From this table, one can compute a histogram representing the relative frequency by age, and we call this histogram as a population age distribution. This kind of population age distribution data is often used in the literature of distributional data analysis \citep{delicado2011dimensionality,
hron2016simplicial, bigot2017geodesic,del2019robust}.
In this study, we use population age distributions for men and women in the 42 districts of Upper and Lower Austria for the year 2020. 
We remove the data for people that are more than $100$ years old, which implies each distribution is supported on the interval $\Omega = [0, 100]$.
Figure \ref{fig:austria_2020_densities} plots their densities obtained by smoothing the histograms.
We see that on average the women distributions have heavier tails on the age range $(70, 100)$ than the men distributions.
We also see that the patterns of variabilities are different between the men and women distributions.
For example, the group of the men distributions has a slightly larger variability on the range $(20, 40)$ than that of the women distributions, and the group of the women distributions has a slightly larger variability than that of the men distributions on the range $(0, 20)$. 
Our goal is to classify these $84 (=42 \times 2)$ distributions into two 
groups with several clustering methods and determine whether the clustering results reflect gender differences.

\subsection{Setup and Result}

We compare the results of the six clustering methods used in the simulation study. 
For selecting the dimension $M$ of geodesic modes of proposed method, we set the threshold value $\tau=0.8$.
By this criterion, $M=1$ is selected. 
We use a computational algorithm developed by \cite{campbell2022efficient} for implementations of convex and geodesic PCA. 
The correct classification rate (cRate) and adjusted Rand index (aRand) are used  as measures of cluster quality. 

\begin{figure*}
    \centering
    \includegraphics[width=99mm]{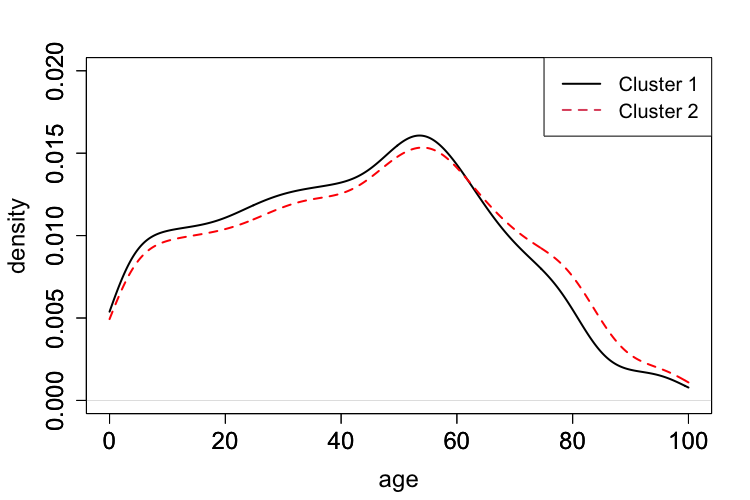}
    \caption{Densities of the estimated Fr\'{eh}cet means of the two clusters. The black solid line is for cluster $1$ (the males group), and the red dashed line is for cluster 2 (the females group).}
    \label{fig:austria_means}
\end{figure*}

\begin{figure*}
    \centering
    \includegraphics[width=139mm]{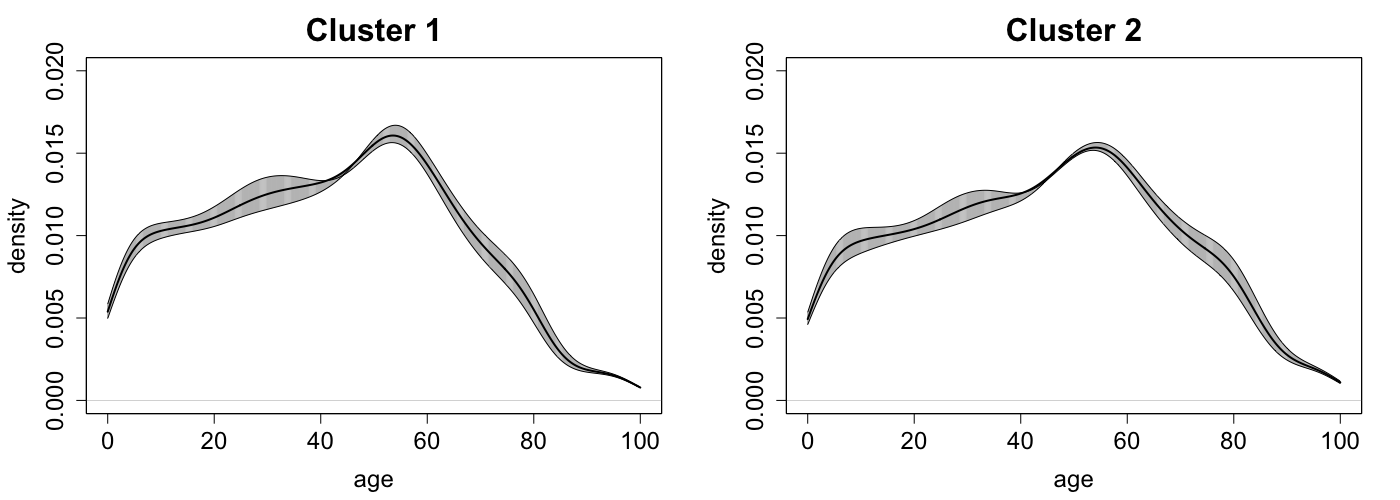}
    \caption{Illustrations of the modes of variations of the two clusters based on the estimated principal geodesics. For each cluster $c \in \{1, 2\}$ and value $\alpha \in \mathbb{R}$, we define a distribution $\hat{\nu}^{(c)}(\alpha) \in \mathcal{P}(\Omega)$ by $\tilde{\nu}^{(c)}(\alpha) = \text{Exp}_{\mu_\ast}(\hat{g}^{(c)} + \alpha \sqrt{\hat{\lambda}_1^{(c)}} \hat{\phi}_1^{(c)})$. Here, $\mu_\ast \in \mathcal{P}(\Omega)$ is some reference measure, $\hat{g}^{(c)} \in \mathcal{L}_{\mu_\ast}^2(\Omega)$ is the estimated mean in the tangent space, $\hat{\phi}_1^{(c)} \in \mathcal{L}_{\mu_\ast}^2(\Omega)$ is the estimated first convex principal direction and $\hat{\lambda}_1^{(c)}$ is the variance of the first convex principal component scores. For each cluster, the black solid line is the density corresponding to $\alpha = 0$, while the boundaries of the shaded region are densities corresponding to $\alpha = \pm 1$. }
    \label{fig:austria_cov}
\end{figure*}

Table \ref{tab:austria_2020_result} summarizes the results. We see that the proposed method outperforms the other methods in distinguishing the gender groups. We further investigate the structure of each cluster by visualizing the estimated Fr\'{e}chet mean and mode of variation.
Figure \ref{fig:austria_means} shows the estimated Fr\'{e}chet means of the two clusters.
We see that the Fr\'{e}chet mean of cluster $2$ (the females group) has a heavier tail on the age range $(60, 100)$ than that of cluster 1 (the males group). This result reasonably reflects the fact that female persons generally live longer than male persons. Figure \ref{fig:austria_cov} illustrates the modes of variations of the two clusters based on the estimated principal geodesics. For cluster 1 (the males group), we see that the largest variability occurs in the range $(20, 40)$ and a large variability occurs in the range $(70, 80)$. For cluster 2 (the female groups), we see that two large variabilities occur in the ranges $(10, 40)$ and $(70, 80)$. 
To see whether the obtained clusters are significantly different, we perform the Komogorov-Smirnov test of two samples on the first convex principal component scores of the data. The $p$-value based on this test is $0.0045$, which provides strong evidence against the equality of the two clusters.
In summary, the proposed method identifies the gender groups with the highest cluster quality, and also provides a visual insight into the clusters. In addition, the obtained clusters are significantly different. 

\begin{table*}[]
    \centering
    \begin{tabular}{ccccccc}
              & CPCA & kCDC & WkM  & $\text{WkM}_{0.01}$ & $\text{WkM}_{0.05}$ & $\text{WkM}_{0.1}$ \\ \hline
        cRate & 0.797 & 0.928 & 0.797 & 0.809 & 0.797 & 0.797 \\
        aRand &0.346 & 0.731  &0.346  &0.375  &0.346&0.346  \\
    \end{tabular}
    \caption{Cluster qualities for the population age distribution data}
    \label{tab:austria_2020_result}
\end{table*}

\section{Discussion}\label{sec:discuss}
In this paper, we propose a novel clustering method for distributional data on the real line. This method performs clustering for distributional data in the spirit of the $k$-centers clustering approach of \cite{chiou2007functional} for functional data. We use geodesic PCA in the Wasserstein space to define the geodesic modes of the variation of clusters, which is used in the reclassification step for determining the cluster membership of each distribution. In contrast to conventional clustering methods like the $k$-means type clustering methods, the proposed method takes account of the differences in both the means and  modes of variation of clusters, potentially improving cluster quality. Our theory demonstrates the validity of our clustering method, which we illustrate through a simulation study and real data analysis.

We note that the proposed clustering method lacks a theoretical support of consistency. For the classical $k$-means clustering, where the cluster centers are only the cluster means, \cite{pollard1981strong} established the almost sure convergence of the set of cluster centers. However, for the $k$-centers type clustering, where the cluster centers include both the cluster means and principal component directions, such consistency results do not seem to have been established yet, and further technical development is needed.

There are several directions for future extensions.
First, one can consider a clustering method for distributional data in the spirit of the correlation-based functional clustering of \cite{chiou2008correlation} that groups curves with similar shapes.
Second, we can consider applying clustering methods to multivariate distributional data. The Wasserstein space of probability distributions on a multi-dimensional Euclidean space can be endowed with the basic concepts of Riemannian manifolds, such as the tangent space, the exponential map, and the logarithmic map (see Section D.1 in the supplementary material for details). However, in the multivariate setting, the logarithmic map generally does not possess the isometric property. This implies that the clear connection between geodesic PCA in the Wasserstein space and convex PCA in the tangent space is not available. Additionally, the Wasserstein distance and logarithmic map generally do not have closed-form expressions in the multivariate setting, which brings about difficulties in numerical computation. These factors need to be considered when extending the proposed clustering method to the multivariate setting. As a special case, for multivariate Gaussian distributions, the Wasserstein distance and the logarithmic map have closed-form expressions. Based on this fact, we present a $k$-centers clustering method for multivariate Gaussian distributions in Section D.2 in the supplementary material.

\section*{Acknowledgements}\,
The authors thank Takeru Matsuda, Yoshikazu Terada and Michio Yamamoto for valuable discussions about the subject of this paper. 
The authors are also grateful to the associate editor and two anonymous referees for their constructive comments which have significantly improved the original version of the paper.
R.Okano was supported by JSPS Grant-in-Aid for JSPS Fellows (22KJ1067). M.Imaizumi was supported by JSPS KAKENHI (21K11780), JST CREST (JPMJCR21D2), and JST FOREST (JPMJFR216I).

\clearpage

\setcounter{section}{0}
\renewcommand{\thesection}{\Alph{section}}

\begin{center}
{\LARGE
Supplementary Material to ``Wasserstein $k$-Centers Clustering for Distributional Data''
}
\end{center}

The supplementary material consists of four sections. Section \ref{sec:cpca} provides a review of convex PCA and its connection to geodesic PCA in the Wasserstein space. Section \ref{sec:implementation} provides implementation details of the proposed method. Section \ref{sec:proof} provides proofs and derivations of our theoretical results. Finally, Section \ref{sec:Mult_Gauss} describes the $k$-centers clustering method for multivariate Gaussian distributions.

\section{Convex Principal Component Analysis}\label{sec:cpca}
\subsection{Formulation}
Convex PCA was originally introduced by \cite{bigot2017geodesic} to analyze geodesic PCA in the Wasserstein space. In this section, we review a general formulation of convex PCA and its link to geodesic PCA in the Wasserstein space.
Let $H$ be a separable Hilbert space over $\mathbb{R}$, with an inner product $\langle \cdot, \cdot \rangle_H$ and norm $\|\cdot \|_H$, and $X$ be a nonempty compact convex subset of $H$. Let $\mathbf{x}$ be an $X$-valued random variable, assumed to be square-integrable in the sense that $\mathbb{E}[\|\mathbf{x}\|_H^2] < \infty$. The mean of $\mathbf{x}$ is defined as the unique element in $\argmin_{y \in X}\mathbb{E}[\|\mathbf{x} - y\|_H^2]$, and we denote it as $\overline{x}$.

To formulate convex PCA, we introduce the following.
For any set $A \subset H$, its dimension, $\text{dim}(A)$, is defined as the dimension of the smallest affine subspace of $H$ containing $A$. For any integer $j \ge 1$, we denote by $\mathcal{C}_{j}$ the family of nonempty, closed and convex subsets $C \subset X$, such that $\text{dim}(C) \le j$ and $\overline{x} \in C$.
For any $x, y \in H$ and nonempty $E \subset H$, we define the distances $d(x, y) = \|x-y\|_H$ and $d(x, E) = \inf_{z \in E}d(x, z)$.

Now we state nested convex PCA. Let $C_0 = \{\overline{x}\}$. For a given integer $M \ge 1$, we define a sequence of convex sets $\{C_j\}_{j=1}^M$ as a solution of a convex PCA problem 
\begin{equation}
            C_j \in \argmin_{C \in \mathcal{C}_{j}, C \supset C_{j - 1}} \mathbb{E}[d^2(\textbf{x}, C)],
            \label{eq:principal_convex}
\end{equation}
for $j=1, ..., M$. The set $C_M$ is called an $(M, \overline{x})$-nested principal convex component of the random element $\textbf{x}$ (Definition 3.3, \cite{bigot2017geodesic}), which represents the mode of variation of $\textbf{x}$. 

In the current setting,  the existence of an $(M, \overline{x})$-nested principal convex component of $\textbf{x}$ is guaranteed for any $M \ge 1$ \citep[Theorem 3.1,][]{bigot2017geodesic}. Furthermore, a solution of the nested convex PCA problem can be obtained by constructing a sequence of orthogonal vectors in $H$ as follows. For $\phi \in H$, we let
\begin{equation}
    V(\phi) = \mathbb{E}\left[\min_{z \in (\overline{x} + \text{span}\{\phi\}) \cap X}\|\textbf{x} - z \|_H^2\right].
\end{equation}
Define 
\begin{equation}
    \phi_1^\ast \in \argmin_{\phi \in H: \|\phi\|_H = 1} V(\phi), 
\end{equation}
and for $j  \ge 2$,
\begin{equation}
    \phi_j^\ast \in \argmin_{\phi \in H: \phi \in P_{j-1}^\top, \|\phi\|_H=1} V(\phi), 
\end{equation}
where $P_{j-1} = \text{span}\{\phi_1, ..., \phi_{j-1}\}$. If we set 
\begin{equation}
     C_j = (\overline{x} + \text{span}\{\phi_1^\ast, ..., \phi_j^\ast\}) \cap X,
     \label{eq:pcc_form}
\end{equation}
 then the set $C_j$ satisfies \eqref{eq:principal_convex} for $j=1, ..., M$.  
\citep[Proposition 3.4,][]{bigot2017geodesic}. 
Hence, $C_M = (\overline{x} + \text{span}\{\phi_1^\ast, ..., \phi_M^\ast\}) \cap X$
is an $(M, \overline{x})$-nested principal convex component of $\textbf{x}$. 
We call the vector $\phi_j^\ast$ as the $j$-th convex principal direction of $\mathbf{x}$.

\subsection{Low-Dimensional Representation}
\label{sec:dimreduc_expvari}
Based on the result of the nested convex PCA problem, a notion of low-dimensional representation or dimension reduction of the random variable $\mathbf{x}$ can be defined. 
Given an $(M, \overline{x})$-nested principal convex component $C_M \subset X$ of $\mathbf{x}$, we define the $M$-dimensional representation of $\mathbf{x}$ as 
\begin{equation}
    \tilde{\textbf{x}} =  \argmin_{z \in C_M} \|\mathbf{x} - z\|_H^2,
    \label{eq:Mdim_rep}
\end{equation}
which uniquely exists by the Hilbert projection theorem. Especially, if $C_M$  has the form  
$C_M = (\overline{x} + \text{span}\{\phi_1^\ast, ..., \phi_M^\ast\}) \cap X$ with the convex principal directions $\{\phi_j^\ast\}_{j=1}^M$, 
there exist coefficients $\xi_1, ..., \xi_M \in \mathbb{R}$ such that 
\begin{equation}
    \tilde{\textbf{x}}
    = 
    \overline{x} 
    +
    \sum_{j=1}^M \xi_j \phi_j^\ast.
    \label{eq:Mdim_expansion}
\end{equation}
We call the scalar $\xi_j$ as the $j$-th convex principal component score of $\mathbf{x}$.
Note that in contrast to the ordinal PCA, the $M$-dimensional representation $\tilde{\textbf{x}}$ does not always obtained by orthogonally projecting $\mathbf{x}$ onto the space $ \overline{x} + \text{span}\{\phi_1^\ast, ..., \phi_{M}^\ast\}$, and thus
the convex principal component score $\xi_j$ in \eqref{eq:Mdim_expansion} is not necessarily equal to the inner product $\langle \mathbf{x} - \overline{x}, \phi_j^\ast \rangle_H$.

\subsection{Link to Geodesic PCA in the Wasserstein Space}
\label{sec:link_GPCA}

The notion of convex PCA is strongly linked to geodesic PCA in the Wasserstein space. Let $\boldsymbol{\mu}$ be a square-integrable $\mathcal{P}(\Omega)$-valued random element with the Fr\'{e}chet mean $\mu_\oplus$, where $\Omega$ is assumed to be a compact interval in $\mathbb{R}$. 
With an absolutely continuous reference measure $\mu_\ast$, define a random variable $\mathbf{g} = \text{Log}_{\mu_\ast}\boldsymbol{\mu}$, and denote its mean as $\overline{g}$.
Here, the random variable $\mathbf{g}$ takes values in the range of the logarithmic map  $V_{\mu_\ast}(\Omega)$, which is compact and convex in the tangent space $\mathcal{L}_{\mu_\ast}^{2}(\Omega)$. 
Let $C_M \subset V_{\mu_\ast}(\Omega)$ be an $(M, \overline{g})$-nested principal convex component of $\mathbf{g}$ obtained by applying the nested convex PCA to $H = \mathcal{L}_{\mu_\ast}^2(\Omega), X = V_{\mu_\ast}(\Omega)$ and $\mathbf{x} = \mathbf{g}$. If we set $G_M = \text{Exp}_{\mu_\ast}(C_M)$, then the set $G_M$ is an $(M, \mu_{\oplus})$-nested principal geodesic of $\boldsymbol{\mu}$ \citep[Proposition 4.4,][]{bigot2017geodesic}. 
 In Figure 1, we illustrated this relationship between geodesic PCA in the Wasserstein space and convex PCA in the tangent space.

\section{Implementation Details}
\label{sec:implementation}

\subsection{Estimation of Distributional Data}
\label{sec:estimation_distributoin}
Let assume that the $n$ distributional data $\nu_1, ..., \nu_n \in \mathcal{P}(\Omega)$ are not directly observed, and instead we observe for each $i=1, ..., n$, a collection of independent measurements $\{Y_{il}\}_{l=1}^{N_i}$ sampled from $\nu_i$.  Here, $N_i$ is the sample size which may vary across distributions.
In this case, we estimate the distribution $\nu_i$ from $\{Y_{il}\}_{l=1}^{N_i}$ before implementing the clustering method. 
One option is estimating the distribution function $F_i$ of $\nu_i$ with the empirical distribution function $\hat{F}_i$ of $\{Y_{il}\}_{l=1}^{N_i}$,
\begin{equation}
    \hat{F}_i(x)
    =
    \frac{1}{N_i} \sum_{l=1}^{N_i}I_{(-\infty, x]}(Y_{il}).
\end{equation}
The quantile function $F_i^{-1}$ of $\nu_i$ is then estimated by converting $\hat{F}_i$ to a quantile function $\hat{F}_i^{-1}$ by left continuous inversion, 
\begin{equation}
    \hat{F}_i^{-1}(u)
    =
    \inf\{x \in \mathbb{R}: \hat{F}_i(x) \ge u\}.
\end{equation}
Another option is estimating the density function of $\nu_i$ from $\{Y_{il}\}_{l=1}^{N_i}$ and then computing the distribution and quantile functions by integration and inversion. 

\subsection{Selection of Dimension of Geodesic Modes by Convex PCA} 
\label{sec:alternative_select_m}
The notion of explained variation for geodesic PCA in the Wasserstein space described in Section 3.2
is equivalent to that for convex PCA in the tangent space. Specifically, for any  $M' \in \mathbb{N}$, 
let  $\hat{C}_{M'} \subset \mathcal{L}_{\hat{\mu}_\ast}^2(\Omega)$ be an $(M', \overline{g})$-nested principal convex component of the $n$ transformed data $g_i = \text{Log}_{\hat{\mu}_\ast}\nu_i, i=1, ..., n$. 
Then letting $\tilde{g}_i = \argmin_{f \in \hat{C}_{M'}}\|g_i - f\|_{\hat{\mu}_\ast}$, the cumulative proportion of variation explained by 
$\hat{C}_{M'}$ is defined as 
\begin{align}
    EV_c(\hat{C}_{M'}) = \frac{n^{-1}\sum_{i=1}^n\|\tilde{g}_i- \overline{g}\|_{\hat{\mu}_\ast}^2}{TV_c},
\end{align}
where  $TV_c = n^{-1}\sum_{i=1}^n\|{g}_i - \overline{g}\|_{\mu_\ast}^2$ is the total variation. 
For further details of this notion of variation explained, see Section 2 of \cite{campbell2022efficient}.
By the isometric property of the logarithmic map, the cumulative proportion of variation of transformed data $g_1, ..., g_n$ that is explained by the principal convex component $\hat{C}_{M'}$ is equal to that of distributional data $\nu_1, ..., \nu_n$ explained by the principal geodesic $\hat{G}_{M'} = \text{Exp}_{\hat{\mu}_\ast}(\hat{C}_{M'})$, namely,  
\begin{equation}
    EV_c(\hat{C}_{M'}) = EV(\hat{G}_{M'})
\end{equation}
for any $M' \in \mathbb{N}$. We use this fact when calculating cumulative proportions of variation explained by principal geodesics in practice.

\section{Proofs}
\label{sec:proof}
\subsection{Proofs of Propositions 1 and 2}
For the proofs, we define stochastic models of the random element $\mathbf{g}^{(c)}$. 
Under Assumption 1, $m^{(c)}$ and $m^{(d)}$ are the mean functions of $\mathbf{g}^{(c)}$ and $\mathbf{g}^{(d)}$, and $\rho_j^{(c)}$ and $\rho_j^{(d)}$ are the $j$-th convex principal directions of $\mathbf{g}^{(c)}$ and $\mathbf{g}^{(d)}$, respectively.
Hence, $(1, m^{(c)})$ and $(1, m^{(d)})$-nested principal convex components of $\mathbf{g}^{(c)}$ and $\mathbf{g}^{(d)}$ are defined as $C^{(c)} = (m^{(c)} + \text{span}\{\rho_1^{(c)}\}) \cap V_{\mu_\ast}(\Omega)$ and $C^{(d)} = (m^{(d)} + \text{span}\{\rho_1^{(d)})\} \cap V_{\mu_\ast}(\Omega)$, respectively. Then for the random variable $\mathbf{g}^{(c)}$, we define
stochastic models $\tilde{\mathbf{g}}_c^{(c)}$ and $\tilde{\mathbf{g}}_c^{(d)}$ as
\begin{equation}
    \tilde{\mathbf{g}}_c^{(c)} = \Pi_{C^{(c)}}\mathbf{g}^{(c)} \quad \text{and} \quad 
    \tilde{\mathbf{g}}_c^{(d)} = \Pi_{C^{(d)}}\mathbf{g}^{(c)},
\end{equation}
respectively. Under Assumption 1, we have $\tilde{\mathbf{g}}_c^{(c)} = m^{(c)} + \xi_1^{(c)}\rho_1^{(c)}$. Additionally, under Assumptions 
1 and 2, we have $\tilde{\mathbf{g}}_c^{(d)} = m^{(d)} + \langle \mathbf{g}^{(c)} - m^{(d)}, \rho_1^{(d)} \rangle_{\mu_\ast}\rho_1^{(d)}$. 
\begin{proof}[Proof of Proposition 1]
    By the isometric property of the logarithmic map, we have 
    \begin{equation}
        d_W(\boldsymbol{\nu}^{(c)}, \tilde{\boldsymbol{\nu}}_{c}^{(c)})
        =
        \|\mathbf{g}^{(c)} - \tilde{\mathbf{g}}_c^{(c)}\|_{\mu_\ast} \quad
        \text{and}
        \quad 
        d_W(\boldsymbol{\nu}^{(c)}, \tilde{\boldsymbol{\nu}}_{c}^{(d)})
        =
        \|\mathbf{g}^{(c)} - \tilde{\mathbf{g}}_c^{(d)}\|_{\mu_\ast}.
        \label{eq:isometric}
    \end{equation}
Note that under Assumption 1, it holds that $\xi_1^{(c)} = \langle \mathbf{g}^{(c)} -  m^{(c)}, \rho_1^{(c)}\rangle_{\mu_\ast}$. Thus, we have
\begin{align}
    \|\mathbf{g}^{(c)} - \tilde{\mathbf{g}}_c^{(c)}\|_{\mu_\ast}^2
    &=
    \|\mathbf{g}^{(c)} - m^{(c)} -\langle \mathbf{g}^{(c)} -  m^{(c)}, \rho_1^{(c)}\rangle_{\mu_\ast} \rho_1^{(c)}\|_{\mu_\ast}^2 \\ 
    &= \|\mathbf{g}^{(c)} - m^{(c)}\|_{\mu_\ast}^2 - \langle \mathbf{g}^{(c)} -  m^{(c)}, \rho_1^{(c)}\rangle_{\mu_\ast}^2. 
    \label{eq:c_expand}
\end{align}
Additionally, under Assumptions 1 and 2, we have
\begin{align}
    \|\mathbf{g}^{(c)} - \tilde{\mathbf{g}}_c^{(d)}\|_{\mu_\ast}^2
    &=
    \|\mathbf{g}^{(c)} - m^{(d)} -\langle \mathbf{g}^{(c)} -  m^{(d)}, \rho_1^{(d)}\rangle_{\mu_\ast} \rho_1^{(d)}\|_{\mu_\ast}^2 \\ 
    &=
    \|\mathbf{g}^{(c)} - m^{(d)}\|_{\mu_\ast}^2
    -
    \langle \mathbf{g}^{(c)} -  m^{(d)}, \rho_1^{(d)}\rangle_{\mu_\ast}^2. 
    \label{eq:d_expand}
\end{align}
Combing \eqref{eq:isometric}, \eqref{eq:c_expand}, \eqref{eq:d_expand}, under the assumption $m^{(c)} = m^{(d)} = m$, we obtain
\begin{align*}
    &P(d_W(\boldsymbol{\nu}^{(c)}, \tilde{\boldsymbol{\nu}}_{c}^{(c)}) < d_W(\boldsymbol{\nu}^{(c)}, \tilde{\boldsymbol{\nu}}_{c}^{(d)})) \\
    &=
    P(\|\mathbf{g}^{(c)} - \tilde{\mathbf{g}}_c^{(c)}\|_{\mu_\ast} < \|\mathbf{g}^{(c)} - \tilde{\mathbf{g}}_c^{(d)}\|_{\mu_\ast}) \\ 
    &=
    P(\|\mathbf{g}^{(c)} - \tilde{\mathbf{g}}_c^{(c)}\|_{\mu_\ast}^2 < \|\mathbf{g}^{(c)} - \tilde{\mathbf{g}}_c^{(d)}\|_{\mu_\ast}^2) \\ 
    &=
    P(\langle \mathbf{g}^{(c)} -  m, \rho_1^{(c)}\rangle_{\mu_\ast}^2 - \langle \mathbf{g}^{(c)} -  m, \rho_1^{(d)}\rangle_{\mu_\ast}^2 > 0) \\
    &=
    P(\langle \mathbf{g}^{(c)} - m, \rho_1^{(c)} + \rho_1^{(d)} \rangle_{\mu_\ast} > 0, \langle \mathbf{g}^{(c)} - m, \rho_1^{(c)} - \rho_1^{(d)} \rangle_{\mu_\ast} > 0) \\ 
    &\,\,\,\,\,\,+P(\langle \mathbf{g}^{(c)} - m, \rho_1^{(c)} + \rho_1^{(d)} \rangle_{\mu_\ast} < 0, \langle \mathbf{g}^{(c)} - m, \rho_1^{(c)} - \rho_1^{(d)} \rangle_{\mu_\ast} < 0).
\end{align*}
\end{proof}

\begin{proof}[Proof of Proposition 2]
As in the proof of Proposition 1, we have the equations (\ref{eq:isometric}) and $\|\mathbf{g}^{(c)} - \tilde{\mathbf{g}}_c^{(c)}\|_{\mu_\ast}^2$ and 
$\|\mathbf{g}^{(c)} - \tilde{\mathbf{g}}_c^{(d)}\|_{\mu_\ast}^2$ are expressed as (\ref{eq:c_expand}) and (\ref{eq:d_expand}), respectively. For the first and second terms in (\ref{eq:d_expand}), we have 
\begin{align}
    \|\mathbf{g}^{(c)} - m^{(d)}\|_{\mu_\ast}^2
    =
    \|\mathbf{g}^{(c)} - m^{(c)}\|_{\mu_\ast}^2 + \|m^{(c)} - m^{(d)}\|_{\mu_\ast}^2
    +
    2\langle \mathbf{g}^{(c)} - m^{(c)}, m^{(c)} - m^{(d)} \rangle_{\mu_\ast}
\end{align}
and
\begin{align}
    \langle \mathbf{g}^{(c)} - m^{(d)}, \rho_1^{(d)} \rangle_{\mu_\ast}^2
    &= \langle \mathbf{g}^{(c)} - m^{(c)}, \rho_1^{(d)}\rangle_{\mu_\ast}^2
    +
    \langle m^{(c)} - m^{(d)}, \rho_1^{(d)}
    \rangle_{\mu_\ast}^2  \\
    &\,\,\,\,\,\,+
    2\langle \mathbf{g}^{(c)} - m^{(c)}, \rho_1^{(d)} \rangle_{\mu_\ast} \langle 
     m^{(c)} - m^{(d)}, \rho_1^{(d)} \rangle_{\mu_\ast}, 
\end{align}
respectively. Thus, with $\psi = 2m^{(c)} - 2m^{(d)} - 2\langle m^{(c)} - m^{(d)}, \rho_1\rangle_{\mu_\ast}\rho_1$,  we have
\begin{align}
    \|\mathbf{g}^{(c)} - \tilde{\mathbf{g}}_{c}^{(d)} \|_{\mu_\ast}^2
    &=
    \|\mathbf{g}^{(c)} - m^{(c)}\|_{\mu_\ast}^2
    -\langle \mathbf{g}^{(c)} - m^{(d)}, \rho_1^{(d)} \rangle_{\mu_\ast}^2
    +
    \langle \mathbf{g}^{(c)} - m^{(c)}, \psi\rangle_{\mu_\ast} \\ 
    &\,\,\,\,\,\,+
    \|m^{(c)} - m^{(d)}\|_{\mu_\ast}^2
    -
    \langle m^{(c)}-m^{(d)}, \rho_1^{(d)} \rangle_{\mu_\ast}^2. 
    \label{eq:d_expand_2}
\end{align}
Combining \eqref{eq:isometric}, \eqref{eq:c_expand} and \eqref{eq:d_expand_2}, under the assumption $\rho_j^{(c)} = \rho_j^{(d)} = \rho_j$ for $j=1, ..., J$, we obtain
\begin{align*}
    &P(d_W(\boldsymbol{\nu}^{(c)}, \tilde{\boldsymbol{\nu}}_{c}^{(c)}) < d_W(\boldsymbol{\nu}^{(c)}, \tilde{\boldsymbol{\nu}}_{c}^{(d)})) \\
    &=
    P(\|\mathbf{g}^{(c)} - \tilde{\mathbf{g}}_c^{(c)}\|_{\mu_\ast} < \|\mathbf{g}^{(c)} - \tilde{\mathbf{g}}_c^{(d)}\|_{\mu_\ast}) \\ 
    &=
    P(\|\mathbf{g}^{(c)} - \tilde{\mathbf{g}}_c^{(c)}\|_{\mu_\ast}^2 < \|\mathbf{g}^{(c)} - \tilde{\mathbf{g}}_c^{(d)}\|_{\mu_\ast}^2) \\ 
    &=
    P(\langle \mathbf{g}^{(c)} - m^{(c)}, \psi \rangle_{\mu_\ast} > \langle m^{(c)} - m^{(d)}, \rho_1 \rangle_{\mu_\ast}^2 - \|m^{(c)} - m^{(d)}\|_{\mu_\ast}^2).
\end{align*}
\end{proof}

\subsection{Derivations of Results in Remarks  7 and 8}
We first derive the result in Remark 7. Let assume the $J$-dimensional random vector $(\xi_1^{(c)}, ..., \xi_J^{(c)})$ follows the zero-mean Gaussian distribution with covariance matrix (8). Then with a two-dimensional zero-mean Gaussian random vector $(W_1, W_2)$ with covariance matrix 
        \begin{equation}
        \begin{pmatrix}
                \sum_{j=1}^J \mathrm{Var}(\xi_j^{(c)})\langle \rho_j^{(c)}, \rho_1^{(c)} + \rho_1^{(d)} \rangle_{\mu_\ast}^2 & \sum_{j=1}^J \mathrm{Var}(\xi_j^{(c)}) \langle \rho_j^{(c)}, \rho_1^{(c)} + \rho_1^{(d)} \rangle_{\mu_\ast}
                \langle \rho_j^{(c)}, \rho_1^{(c)} - \rho_1^{(d)} \rangle_{\mu_\ast} \\ 
                \sum_{j=1}^J \mathrm{Var}(\xi_j^{(c)}) \langle \rho_j^{(c)}, \rho_1^{(c)} + \rho_1^{(d)} \rangle_{\mu_\ast}
                \langle \rho_j^{(c)}, \rho_1^{(c)} - \rho_1^{(d)} \rangle_{\mu_\ast} & 
                \sum_{j=1}^J \mathrm{Var}(\xi_j^{(c)})\langle \rho_j^{(c)}, \rho_1^{(c)} - \rho_1^{(d)} \rangle_{\mu_\ast}^2
       \end{pmatrix},
    \end{equation}
        the probability (7) is expressed as $\mathbb{P}(W_1 > 0, W_2 > 0)
            +
            \mathbb{P}(W_1 < 0, W_2 < 0)$, which is equal to $0.5 + \pi^{-1}\arcsin{\mathrm{Corr}(W_1, W_2)}$ (see Section 15.10 in \cite{stuart2010kendall} for this equality). 
            Furthermore, 
        assuming $\mathrm{span}\{\rho_1^{(c)}, ..., \rho_J^{(c)}\} = \mathrm{span}\{\rho_1^{(d)}, ..., \rho_J^{(d)}\}$, there exists an index $\ell \in \{1, ..., J\}$ such that $\rho_\ell^{(c)} = \rho_1^{(d)}$ and $\rho_j^{(c)} \neq \rho_1^{(d)}$ for $j \neq \ell$. If $\ell \ge 2$ we have 
        \begin{gather*}
            \sum_{j=1}^J \mathrm{Var}(\xi_j^{(c)})\langle \rho_j^{(c)}, \rho_1^{(c)} + \rho_1^{(d)} \rangle_{\mu_\ast}^2 = \mathrm{Var}(\xi_1^{(c)}) + \mathrm{Var}(\xi_\ell^{(c)}), \\ 
             \sum_{j=1}^J \mathrm{Var}(\xi_j^{(c)}) \langle \rho_j^{(c)}, \rho_1^{(c)} + \rho_1^{(d)} \rangle_{\mu_\ast}
                \langle \rho_j^{(c)}, \rho_1^{(c)} - \rho_1^{(d)} \rangle_{\mu_\ast} = \mathrm{Var}(\xi_1^{(c)}) - \mathrm{Var}(\xi_\ell^{(c)}), \\
            \sum_{j=1}^J \mathrm{Var}(\xi_j^{(c)})\langle \rho_j^{(c)}, \rho_1^{(c)} - \rho_1^{(d)} \rangle_{\mu_\ast}^2 = \mathrm{Var}(\xi_1^{(c)}) + \mathrm{Var}(\xi_\ell^{(c)}).
        \end{gather*}
        This implies the correlation between $W_1$ and $W_2$ is given by $\{\mathrm{Var}(\xi_1^{(c)}) - \mathrm{Var}(\xi_\ell^{(c)})\}/\{\mathrm{Var}(\xi_1^{(c)}) + \mathrm{Var}(\xi_\ell^{(c)})\}$, and the probability (7) is eventually expressed as (9). 

Next, we derive the result in Remark 8. Let assume the random vector $(\xi_1^{(c)}, ..., \xi_J^{(c)})$ follows the zero-mean Gaussian distribution with covariance matrix (8).
Then with a zero-mean Gaussian random variable $V$ with variance $\sum_{j=1}^J \mathrm{Var}(\xi_j^{(c)}) \langle \rho_j, \psi \rangle_{\mu_\ast}^2$, the probability (10) is expressed as $\mathbb{P}(V > \langle m^{(c)} - m^{(d)}, \rho_1 \rangle_{\mu_\ast}^2 - \|m^{(c)} - m^{(d)}\|_{\mu_\ast}^2)$. Suppose $J \ge 2$. 
 If $m^{(c)} - m^{(d)} \in \mathrm{span}\{\rho_2, ..., \rho_J\}$, then the variance of $V$ is bounded above by $4\mathrm{Var}(\xi_1^{(c)})\|m^{(c)} - m^{(d)}\|_{\mu_\ast}^2$, and $\langle m^{(c)} - m^{(d)}, \rho_1 \rangle_{\mu_\ast}^2 - \|m^{(c)} - m^{(d)}\|_{\mu_\ast}^2 $ reduces to $ -\|m^{(c)} - m^{(d)}\|_{\mu_\ast}^2$. This implies the probability (10) is bounded below by (11).

\section{$k$-Centers Clustering for Multivariate Gaussian  Distributions}\label{sec:Mult_Gauss}
\subsection{Optimal Transport of Gaussian Distributions on $\mathbb{R}^d$}\label{sec:Gauss_OT}
In this subsection, we provide some background on optimal transport of Gaussian distributions on $\mathbb{R}^d$, for $d \ge 1$. Firstly, we explain  optimal transport of general probability distributions on $\mathbb{R}^d$ \citep{ambrosio2008gradient, villani2008optimal, panaretos2020invitation}. Let $\mathcal{P}_2(\mathbb{R}^d)$ be the set of Borel probability measures on $\mathbb{R}^d$ with finite second moments.
The 2-Wasserstein distance between $\mu_1, \mu_2 \in \mathcal{P}_2(\mathbb{R}^d)$ is defined by 
\begin{equation}
    d_W(\mu_1, \mu_2)
    =
    \left(\inf_{\pi \in \Gamma(\mu_1, \mu_2)} \int_{\mathbb{R}^d \times \mathbb{R}^d} \|x-y\|^2 d\pi(x, y)\right)^{1/2},
\end{equation}
where $\|\cdot\|$ denotes the Euclidean norm on $\mathbb{R}^d$, and the infimum is taken over the set $\Gamma(\mu_1, \mu_2)$ of all couplings of $\mu_1$ and $\mu_2$. The 2-Wasserstein distance $d_W$ is a metric on $\mathcal{P}_2(\mathbb{R}^d)$, and the metric space $(\mathcal{P}_2(\mathbb{R}^d), d_W)$ is called the Wasserstein space. For two given measures $\mu_\ast, \mu \in \mathcal{P}_2(\mathbb{R}^d)$, any map $\mathbf{t}: \mathbb{R}^d \to \mathbb{R}^d$ that minimizes Monge's problem $\min_{\mathbf{t}\# \mu_\ast = \mu}\int_{\mathbb{R}^d}\|\mathbf{t}(x) - x\|^2 d\mu_\ast(x)$ is called an optimal transport map from $\mu_\ast$ to $\mu$. If $\mu_\ast$ is absolutely continuous with respect to Lebesgue measure on $\mathbb{R}^d$, such optimal transport map uniquely exists, and we denote it as $\mathbf{t}_{\mu_\ast}^\mu$. We note that when $d=1$, the optimal transport map is given by $\mathbf{t}_{\mu_\ast}^\mu = F^{-1} \circ F_\ast$, where $F_\ast$ and $F^{-1}$ are the distribution function of $\mu_\ast$ and the quantile function of $\mu$, respectively. Analogously to the case of $d=1$, the notions of tangent space, exponential map and logarithmic map at $\mu_\ast$ can be defined. Specifically, the tangent space is the Hilbert space $\mathcal{L}_{\mu_\ast}^2(\mathbb{R}^d)$ of $\mathbb{R}^d$-valued functions $g$ on $\mathbb{R}^d$ that are $\mu_\ast$-square-integrable in the sense $\int_{\mathbb{R}^d}\|g\|^2 d\mu_\ast < \infty$. This space $\mathcal{L}_{\mu_\ast}^2(\mathbb{R}^d)$ is equipped with an inner product $\langle \cdot, \cdot \rangle_{\mu_\ast}$ defined by $\langle g_1, g_2 \rangle_{\mu_\ast} = \int_{\mathbb{R}^d} g_1^\top g_2 d\mu_\ast$ and norm $\|\cdot\|_{\mu_\ast}$ defined by $\|g\|_{\mu_\ast} = \langle g, g \rangle_{\mu_\ast}^{1/2}$.
The exponential map $\text{Exp}_{\mu_\ast}: \mathcal{L}_{\mu_\ast}^2(\mathbb{R}^d) \to \mathcal{P}_2(\mathbb{R}^d)$ is defined by $\text{Exp}_{\mu_\ast}g = (g + \text{id})\#\mu_\ast$, and the logarithmic map $\text{Log}_{\mu_\ast}: \mathcal{P}_2(\mathbb{R}^d) \to \mathcal{L}_{\mu_\ast}^2(\mathbb{R}^d)$ is defined by $\text{Log}_{\mu_\ast}\mu = \mathbf{t}_{\mu_\ast}^\mu - \text{id}$.

Next, we explain optimal transport of multivariate Gaussian measures (see Section 1.6.3 of \cite{panaretos2020invitation} for further details). We simplify the setting by focusing on centered Gaussian measures. Let $\mathcal{G}_0(d)$ be the set of Gaussian measures on $\mathbb{R}^d$ with zero means. For Gaussian measures $\mu_1 = N(0, \Sigma_1)$ and $ \mu_2 = N(0, \Sigma_2)$ in 
$\mathcal{G}_0(d)$, the Wasserstein distance between them  is explicitly expressed as 
\begin{equation}
    d_W(\mu_1, \mu_2)
    =
    \sqrt{\text{tr}[\Sigma_1 + \Sigma_2  - 2(\Sigma_1^{1/2}\Sigma_2\Sigma_1^{1/2})^{1/2}]}.
\end{equation}
Here, for any positive semidefinite and symmetric matrix $\Sigma$, we denote its principal square root as $\Sigma^{1/2}$.
Also, for two given Gaussian measures $\mu_\ast = N(0, \Sigma_\ast)$ and $\mu = N(0, \Sigma)$ in $\mathcal{G}_0(d)$, where $\mu_\ast$ is assumed to be non-singular, the optimal transport map from $\mu_\ast$ to $\mu$ is expressed as $\mathbf{t}_{\mu_\ast}^{\mu}(x) = \Sigma_{\ast}^{-1/2}[\Sigma_\ast^{1/2}\Sigma \Sigma_\ast^{1/2}]^{1/2}\Sigma_\ast^{-1/2}x$. 

Now, we consider the notions of tangent space, exponential and logarithmic maps of the space of Gaussian distributions $(\mathcal{G}_0(d), d_W)$ at the non-singular reference measure $\mu_\ast = N(0, \Sigma_\ast) \in \mathcal{G}_0(d)$. Let $\text{Sym}(d)$ denote the set of  $d \times d$ symmetric matrices, and denote with $g_V$ the linear transformation on $\mathbb{R}^d$ represented by a matrix $V \in \text{Sym}(d)$. Then, for any $\mu = N(0, \Sigma)$ in $\mathcal{G}_0(d)$, we have $\text{Log}_{\mu_\ast}\mu = \mathbf{t}_{\mu_\ast}^{\mu}-\text{id} = g_{\Sigma_{\ast}^{-1/2}[\Sigma_\ast^{1/2}\Sigma \Sigma_\ast^{1/2}]^{1/2}\Sigma_\ast^{-1/2}} - g_{I_d} = g_{\Sigma_{\ast}^{-1/2}[\Sigma_\ast^{1/2}\Sigma \Sigma_\ast^{1/2}]^{1/2}\Sigma_\ast^{-1/2} - I_d}$, where $I_d$ denotes the identity matrix of size $d$. Moreover, we have $\langle g_{V_1}, g_{V_2} \rangle_{\mu_\ast} = \int_{\mathbb{R}^d} (V_1x)^{\top}(V_2x)d\mu_\ast(x) = \text{tr}[V_1 \Sigma_\ast V_2]$ for any $V_1, V_2 \in \text{Sym}(d)$, and $\text{Exp}_{\mu_\ast}g_V = (g_V + \text{id})\# \mu_\ast = N(0, (V+I_d)\Sigma_\ast (V+I_d))$ for any $V \in \text{Sym}(d)$. Based on these observations and the identification of the linear transformation $g_V$ with the matrix $V$, we again define the tangent space, exponential and logarithmic maps of the space $(\mathcal{G}_0, d_W)$ as follows. The tangent space at the reference measure $\mu_\ast = N(0, \Sigma_\ast)$ is defined as a finite-dimensional inner product space $(\text{Sym}(d), \langle \cdot, \cdot \rangle_{\Sigma_\ast})$, where 
$\langle \cdot, \cdot \rangle_{\Sigma_\ast}$ is an inner product defined by $\langle V_1, V_2 \rangle_{\Sigma_\ast} = \text{tr}(V_1 \Sigma_\ast V_2)$. 
We denote the norm induced by this inner product as $\|\cdot\|_{\Sigma_\ast}$.
The exponential map $\text{Exp}_{\Sigma_\ast}: \text{Sym}(d) \to \mathcal{G}_0(d)$ is then defined by 
\begin{equation}
    \text{Exp}_{\Sigma_\ast} V
    =
    N(0, (V + I_d)\Sigma_\ast(V+I_d)), 
\end{equation}
and the logarithmic map $\text{Log}_{\Sigma_\ast}: \mathcal{G}_0(d) \to \text{Sym}(d)$ is defined by 
\begin{equation}
     \text{Log}_{\Sigma_\ast}\mu
     =
     \Sigma_\ast^{-1/2}[\Sigma_\ast^{1/2}\Sigma \Sigma_\ast^{1/2}]^{1/2}\Sigma_\ast^{-1/2} - I_d,
 \end{equation}
 for $\mu = N(0, \Sigma) \in \mathcal{G}_0(d)$. We denote the range of the logarithmic map $\text{Log}_{\Sigma_\ast}$ as 
$V_{\Sigma_\ast}(d) \subset \text{Sym}(d)$. 
It can be easily checked that $V_{\Sigma_\ast}(d)$ is expressed as 
\begin{equation}
    V_{\Sigma_\ast}(d)
    =
    \{V \in \text{Sym}(d): \text{$V + I_d$ is positive semidefinite} \},
\end{equation}
and hence $V_{\Sigma_\ast}(d)$ is closed and convex in $\text{Sym}(d)$.

In contrast to the one-dimensional case, the space $(\mathcal{G}_0(d), d_W)$ is not isometric to the set $V_{\Sigma_\ast}(d)$ in $\text{Sym}(d)$. The following proposition shows that the logarithmic map $\text{Log}_{\Sigma_\ast}$ has an isometric property on a specific class of Gaussian distributions. 

\begin{prp}[cf. \cite{panaretos2020invitation}, Section 2.3.2]
Let $\mathscr{C} \subset \mathcal{G}_0(d)$ be a class of Gaussian distributions such that $\mu_\ast \in \mathscr{C}$ and $\Sigma_1\Sigma_2 = \Sigma_2\Sigma_1$ for any $\mu_1 = N(0, \Sigma_1), \mu_2 = N(0, \Sigma_2)$ in $\mathscr{C}$. Then we have 
\begin{equation}
    d_W(\mu_1, \mu_2)
    =
    \|\mathrm{Log}_{\Sigma_\ast}\mu_1 - \mathrm{Log}_{\Sigma_\ast}\mu_2\|_{\Sigma_\ast}
\end{equation}
for any $\mu_1, \mu_2 \in \mathscr{C}$.
\end{prp}

\subsection{Clustering Procedure}
\paragraph{Setting}
We propose a clustering method for multivariate Gaussian distributions based on the $k$-centers clustering approach.
Suppose there are $n$ Gaussian distributions $\nu_1 = N(0, \Sigma_1), ..., \nu_n = N(0, \Sigma_n)$ in $\mathcal{G}_0(d)$, and we aim to classify them into $K$ groups. As with the case of univariate distributions, the distributions $\nu_i$ may be not directly observed, and instead we observe a collection of independent measurements $\{Y_{il}\}_{l=1}^{N_i}$ sampled from $\nu_i$. In this case, we need to estimate the Gaussian distribution $\nu_i$, especially its covariance matrix $\Sigma_i$, from the measurements. A natural estimator is the empirical covariance matrix
\begin{equation}
    \hat{\Sigma}_i
    =
    \frac{1}{N_i}\sum_{l=1}^{N_i} (Y_{il}-\overline{Y}_i)
    (Y_{il}-\overline{Y}_i)^\top,
\end{equation}
where $\overline{Y}_i = N_i^{-1}\sum_{l=1}^{N_i}Y_{il}$ is the empirical mean. To keep notations simple, we will use the same notation $\nu_i$ for the estimated distribution. The procedure of the proposed clustering method is as follows.

\paragraph{Initial Clustering}
As with the case of univariate distributions, we initially classify the $n$ distributions into $K$ groups by performing dimension reduction based on the stochastic structure of overall data and then applying a conventional clustering algorithm on the resulting low-dimensional vectors.
 Let $\hat{\mu}_\ast \in \mathcal{G}_0(d)$ be a reference Gaussian distribution with a covariance matrix $\hat{\Sigma}_\ast$ and transform the Gaussian distributions $\nu_i$ as $V_i = \text{Log}_{\hat{\Sigma}_\ast}\nu_i, i=1, ..., n$. A typical choice of $\hat{\mu}_\ast$ is the empirical Fr\'{e}chet mean of $\nu_1, ..., \nu_n$ in the space $(\mathcal{G}_0(d), d_W)$,
\begin{equation}
    \hat{\nu}_{\oplus}
    =
    \argmin_{\mu \in \mathcal{G}_0(d)}\frac{1}{n}\sum_{i=1}^n d_W^2(\nu_i, \mu),
\end{equation}
which can be computed efficiently by using the steepest descent algorithm \citep{zemel2019frechet}. 
Let $M$ be an integer such that $1 \le M \le d(d+1)/2$. We have the $M$ convex principal directions $\{\hat{\Phi}^\ast_1, ..., \hat{\Phi}^\ast_M\} \subset \text{Sym}(d)$ by applying convex PCA to the data $V_1, ..., V_n$ with the Hilbert space $( \text{Sym}(d), \langle \cdot, \cdot \rangle_{\Sigma_\ast})$ and convex set $V_{\hat{\Sigma}_\ast}(d)$. 
Then the set $\hat{C}_M = (\overline{V} + \text{span}\{\hat{\Phi}_1, ..., \hat{\Phi}_M\}) \cap V_{\hat{\Sigma}_\ast}(d)$ is an $(M, \overline{V})$-principal convex component of $V_1, ..., V_n$. 
Based on this result, for each $i=1, ..., n$, we obtain the convex principal component scores of $V_i$, $\hat{\xi}_i = (\hat{\xi}_{i1}, ..., \hat{\xi}_{iM}) \in \mathbb{R}^M$.
The initial clustering membership is determined by applying a conventional algorithm such as the $k$-means method to the $M$-dimensional vectors $\hat{\xi}_{i}, i=1, ..., n$. As with the case of univariate distributions, we use the notion of explained variation for convex PCA to choose the dimension $M$ of the principal convex component.

\paragraph{Reclassification}
With the initial clustering results, we use the mean and covariance structures of the clusters to reclassify each data into a best predicted cluster.
Specifically, let  $h_i{(t)} \in \{1, ..., K\}$ be the label of cluster membership for the $i$-th distribution at the $t$-th iteration. 
Given the set of clustering results $\mathcal{H}{(t)} = \{h_i(t): i=1, ..., n\}$, we obtain for each individual $i$ and cluster $c$ the estimates of mean $\overline{V}^{(c)}$ and an $(M, \overline{V}^{(c)})$-nested principal convex component $\hat{C}_{M}^{(c)}$ based on the matrices $V_k$ with $h_k{(t)}=c$ for all $k \neq i$, leaving out the $i$-th matrix. 
Here, $M$ is the value of dimension that was chosen in the initial clustering step.
Given these estimates, we obtain the $i$-th predicted model for each cluster $c$, 
\begin{equation}
    \tilde{V}_{(i)}^{(c)}
    =
    \argmin_{U \in \hat{C}_M^{(c)}} \|V_i - U\|_{\hat{\Sigma}_\ast},
\end{equation}
which is the $M$-dimensional representation of $V_i$ defined as in \eqref{eq:Mdim_rep}. The $i$-th individual is classified into cluster $h_i{(t+1)}$ such that
\begin{equation}
    h_i{(t+1)}
    =
    \argmin_{c \in \{1, ..., K\}}
    \|V_i - \tilde{V}_{(i)}^{(c)} \|_{\hat{\Sigma}_\ast}.
\end{equation}
This step is performed for all $i$, which leads to an updated set of results $\mathcal{H}{(t + 1)}=\{h_i{(t+1)}: i=1, ..., n\}$. The updating procedure is iteratively implemented until no more data can be reclassified.

\bibliographystyle{chicago}
\bibliography{ref}

\end{document}